\documentclass[fleqn,usenatbib]{mnras}

\usepackage{newtxtext,newtxmath}
\usepackage[T1]{fontenc}
\usepackage{ae,aecompl}

\usepackage{graphicx}	% Including figure files
\usepackage{amsmath}	% Advanced maths commands
\usepackage{amssymb}	% Extra maths symbols

\title[Kinematics and stellar archaeology of NGC 613]{The nuclear region of NGC 613 -- II. Kinematics and stellar archaeology}

\author[Patr\'icia da Silva et al.]{
Patr\'icia da Silva$^1$ \thanks{\href{mailto:p.silva2201@gmail.com}{p.silva2201@gmail.com}},
R. B. Menezes$^{2}$ \thanks{\href{mailto:roberto.menezes@maua.br}{roberto.menezes@maua.br}},
J. E. Steiner$^1$ \thanks{\href{mailto:joao.steiner@iag.usp.br}{joao.steiner@iag.usp.br}},
Luciano Fraga$^{3}$
\\
$^1$Instituto de Astronomia, Geof\'isica e Ci\^encias Atmosf\'ericas, Departamento de Astronomia, Universidade de S\~ao Paulo, 05508-090, SP, Brazil\\
$^2$Instituto Mau\'a de Tecnologia, Pra\c{c}a Mau\'a 1, 09580-900, S\~ao Caetano do Sul, SP, Brazil \\
$^3$Laborat\'orio Nacional de Astrof\'isica, 37504-364, Itajub\'a, Minas Gerais, MG, Brazil
}
\date{Accepted: 2020 May 26; Received: 2020 February 20}

\pubyear{2020}

\begin{document}
\label{firstpage}
\pagerange{\pageref{firstpage}--\pageref{lastpage}}
\maketitle

\begin{abstract}
In this work, we continue the study of the central region of NGC~613 by \citet{paty}, by analysing the stellar and gas kinematics and the stellar archaeology in optical and near-infrared data cubes. The high spatial resolution of the Gemini Multi-Object Spectrograph (GMOS) data cube allowed the detection, using spectral synthesis methods, of an inner circumnuclear ring, with a radius of $\sim$~1~arcsec, composed of $\sim$~10$^9$-yr stellar populations. Such a ring is located between the nucleus and the circumnuclear ring composed by H~\textsc{ii} regions detected in previous works. Besides that, there is a stellar rotation around the nucleus and the rings follow the same direction of rotation with different velocities. The intensity-weighted average stellar velocity dispersion at the centre is 92~$\pm$~3~km~s$^{-1}$. Three distinct gas outflow components were detected. The direction of the outflow observed with the H~$\alpha$ emission line is compatible with the direction of the previously observed radio jet. The direction of one of the outflows detected in the [O~\textsc{iii}]$\lambda$5007 emission coincides with the axis of the ionization cone. There is no difference regarding the stellar populations and the stellar kinematics along the double stellar emission, probably separated by a dust lane as mentioned in Paper I, confirming that they are part of the same structure.  
\end{abstract} 


%%%%%%%%%%%%%%%%%%%%%%%%%%%%%%%%%%%%%%%%%%%%%%%%%%

%%%%%%%%%%%%%%%%% BODY OF PAPER %%%%%%%%%%%%%%%%%%

\section{Introduction}

Active galactic nuclei (AGNs) interact with the host galaxy through processes of feeding and feedback \citep{reviewfeeeding}. The study of the gas dynamics determines the proportion of such processes in the galaxy nucleus. In addition, the stellar dynamics determines how the galaxy behaves with the presence of such objects. Together with the stellar archaeology, it is possible to infer evolutionary aspects of the host galaxies from the co-evolution of their nuclei and their circumnuclear regions \citep{reviewcoevolution}.

This work is a continuation of \citet{paty}, hereafter Paper~I, and is focused on the study of the stellar and gas kinematics and of the stellar archaeology of the central region of NGC~613, with the aim of determining a dynamic and evolutionary scenario to this galaxy nucleus. 

NGC~613 is an SB(rs)bc galaxy, located at 26~$\pm$~5 Mpc \citep{distancia}. One of its most outstanding morphological features is the bar, whose size is 139~arcsec ($\sim$ 17~kpc; \citealt{kormendy}). The bar interacts with the nuclear region bringing gas from the outer regions to the centre through a nuclear spiral (\citealt{audibert}; Paper I) interfering in the star formation, which seems to take place in bursts and not in a continuous form \citep{alloin}.

As discussed in Paper~I and by \citet{audibert}, the AGN of NGC 613 has emission-line ratios compatible with the ones of low-ionization nuclear emission-line regions (LINERs) and previous works indicate that it is considerably obscured \citep{castangia2013,ASMUS2015}. The bolometric luminosity of the AGN estimated by \citet{daviesbecca} is 1.6 $\times 10^{42}$ erg s$^{-1}$.

As mentioned in Paper~I and in previous works (\citealt{hummel}; \citealt{BOKERIAU,boker};\citealt{falconIAU,falcon613}; Paper I), NGC~613 has a star-forming ring that contains eight H~\textsc{ii} regions. There is no consensus on how those regions were formed. \citet{falconIAU} and \citealt{boker} suggest two scenarios: the "popcorn" and "pearls on string" models. In the first one, the H~\textsc{ii} regions were formed randomly or at the same time. In the second scenario, the star-forming clouds were created from two primordial clouds (in two distinct points on the ring) that bring gas and dust to the ring; the clouds closer to those feeding regions are younger than the ones far from them, due to the ring rotation. According to those authors, the difference between the He~\textsc{i}, Br~$\gamma$, and [Fe~\textsc{ii}]$\lambda$16436 emission supports this last scenario. On the other hand, \citet{mazzuca} did not find any age gradient between the H~\textsc{ii} regions, whose stars are older than 10 Myr, and they affirm that those regions where formed by gravitation instabilities in the Lindblad resonance region. The star-forming rate seems to be higher to the east portion of the ring \citep{miyamoto1}.

Regarding the gas kinematics, optical observations and images of the [O~\textsc{iii}]$\lambda$5007 emission line show that there is a high-velocity outflow, co-spatial with the radio jet \citep{hummel2}. This outflow seems to disturb the ring structure \citep{boker,falcon613}. \citet{miyamoto2} also identified an outflow of molecular gas in the inner 100~pc. The authors claim that this outflow can be quenching the star formation in this region.

Besides that, the molecular gas kinematics, observed with the Atacama Large Millimeter/Submillimeter Array (ALMA), indicates a rotation of gas in the ring. The portion of the ring to the west presents redshifted emission and that to the east, presents blueshifted emission \citep{miyamoto1,miyamoto2,audibert,combes613}. Such a kinematic behaviour is compatible with the one observed from the images of Br~$\gamma$ by \citet{boker}. An outflow was also detected in the CO(3-2) emission \citep{audibert}, aligned with the radio jet observed by \citet{hummel}. A nuclear spiral of molecular gas was observed and seems to be feeding the nucleus, bringing gas and dust from the bar to the centre (\citealt{audibert}, Paper I).

With infrared integral field unit (IFU) data, \citet{batcheldor} determined two values of the stellar velocity dispersion, one using the cross-correlation method, $\sigma$=~99~$\pm$~2~km~s$^{-1}$, and the other using the maximum penalized likelihood method, $\sigma$=~$149_{-22}^{+19}$~km~s$^{-1}$. These values are compatible with  $\sigma$=~126~$\pm$~19~km~s$^{-1}$ \citep{schechter}, obtained using the Fourier quotient technique \citep{sargent}.

In this work, as in Paper~I, the analysis of the central region of NGC~613 was done using data cubes observed with the IFU from the Gemini Multi-Object Spectrograph (GMOS) at the Gemini-South telescope, with the SOAR Integral Field Spectrograph (SIFS) at the SOAR telescope and the public archive data obtained with the Spectrograph for Integral Field Observations in the Near Infrared (SINFONI) at the Very Large Telescope (VLT).

In section \ref{sec_obs}, we briefly describe the observations and processes of reduction and data treatment. In section \ref{sec_gaskinematics}, we present the analysis regarding the gas kinematics. In section \ref{sec_stellararche}, we show the results obtained from the stellar archaeology. Section \ref{secstellarkinematics} presents the stellar kinematics analysis. Sections \ref{sec_discussion} and \ref{sec_conclusion} present the discussion of the results and their main conclusions, respectively. In appendix \ref{sifs_starlighttotal}, we present the results of the stellar archaeology applied to the whole SIFS data cube.

\section{Observations and data reduction}\label{sec_obs}

Paper~I describes all the details of the data and their treatment processes (see section 2 from Paper~I). All data cubes were treated with the same methodologies developed by our group (see \citealt{rob1,rob2,gmostreat} for more details).

The reduction process of the GMOS data was performed in an \textsc{iraf} environment and that of the SIFS data with scripts developed in Interactive Data Language (\textsc{idl}). The SINFONI data were reduced using the \textsc{gasgano} software. 

After the data reduction and the creation of the data cubes, those were treated using scripts also written in \textsc{idl}. The atmospheric differential refraction correction was applied (except the SIFS data cube, which already had this correction in its hardware) and the data cubes were combined into one, in the form of a median. The SINFONI and SIFS data cubes were re-sampled. The sizes of the spatial pixels (spaxels) of the SINFONI and SIFS data cubes, at the end of the treatment procedure, were 0.625 and 0.1~arcsec, respectively. The GMOS data cube remained with the same spaxel size of 0.05~arcsec.

The Butterworth spatial filtering was applied \citep{gwoods} to all data cubes. After that, the fingerprint removal method was performed and, at the end, we also applied the Richardson--Lucy deconvolution \citep{rich,lucy}. The last process was performed with the aim of obtaining a better spatial resolution without interfering in the data signal (for more details on the data-treatment procedure, see \citealt{rob1,rob2,gmostreat}). Paper~I presents the parameters that we used in this last process. As a result, the GMOS data cube has a full width at half-maximum (FWHM) of the point spread function (PSF) equal to 0.68 arcsec, after 10 iterations. The SINFONI data cube, with the same number of iterations, has the FWHM of the PSF equal to 0.60~arcsec. Lastly, the SIFS data cube has the FWHM of the PSF equal to 1.6~arcsec, using six iterations. 

After the data treatment, the spectra of the GMOS, SIFS, and SINFONI data cubes were all passed to the rest frame, using a redshift value of 0.00494 derived from \citet{redshift613}.

As in Paper~I, we also studied in this paper the ALMA data cube observed in the CO(3-2) line. These data were obtained from the ALMA public archive under program 2015.1.00404.S (PI: Combes). It was not necessary to apply any treatment to the data since it has already good spatial resolution and high quality. The size of the spaxels of this data cube is 0.07~arcsec.

\subsection{Spectral synthesis methods}\label{starlight_def}

In order to study the gas emission from the data cubes and also to analyse the stellar archaeology, spectral synthesis was applied, using the \textsc{starlight} software \citep{starlight}, with a base created from the Medium-resolution Isaac Newton Telescope Library of Empirical Spectra (MILES; \citealt{blazquez}) in the optical wavelengths (GMOS and SIFS data cubes). The spectral synthesis is applied to each spectrum of the data cube, and is performed by a linear combination of the base spectra, in order to fit the observed spectra. With that, it is possible to create flux maps of the detected stellar populations, to obtain histograms of the flux fractions of those stellar populations in a determined region of the FOV of the data cube (which may be the whole FOV or a specific area) and to remove the stellar continuum of the data cube to study the gas emission (as done in Paper~I). This last process corresponds to creating a synthetic stellar data cube, using the synthetic spectra obtained from the spectral synthesis, and then by subtracting this data cube from the original one. We obtain a cube with mainly gas emission, which we call gas data cube. In this work, the gas data cube was used to study the gas kinematics.

The penalized pixel fitting method (pPXF; \citealt{cappellari}) was used in the SINFONI data cube. This method consists of a spectral synthesis with the base spectra convolved with Gauss--Hermite functions. The base used covers only the $K$ band (\citealt{starlightinfrared}). In this case, the pPXF was used to obtain a $K$-band gas data cube, to study the stellar kinematics in the optical (SIFS and GMOS data cubes) and in the determination of the stellar velocity dispersion with the SINFONI data cube. For more details, see section \ref{secstellarkinematics}.

In section \ref{sec_stellararche}, we present the results of the stellar archaeology obtained from the spectral synthesis applied to the extracted spectra from different regions of the data cubes (GMOS and SIFS). We characterized the detected stellar populations by obtaining histograms showing the flux fractions associated with each stellar population. In order to estimate the uncertainty of this process, first of all, we determined the medians of the ages of the detected stellar populations in all these regions of the data cubes. Such medians were calculated using graphs of the accumulated flux fractions as a function of the ages of the stellar populations and taking the ages at which the accumulated flux fractions were equal to 50\%. The uncertainties of these medians were obtained with the following Monte Carlo procedure: First, we constructed, for each extracted spectrum, a histogram representing the spectral noise and fitted a Gaussian function to this histogram. After that, we obtained Gaussian distributions of random noise with the same width of the Gaussian fitted to the initial histogram. Such noise distributions were added to the synthetic stellar spectrum provided by the spectral synthesis of the original spectrum. Finally, we applied the spectral synthesis to all the obtained ``noisy'' spectra and calculated the median of the ages of the detected stellar populations. The uncertainty of the median of the ages was taken as the standard deviation of all the median values obtained with this process. The representative uncertainty of the process was taken as the highest value of the uncertainties determined for the median ages. The result obtained is 0.02~dex. 

When we take the results from the spectral synthesis it is important to notice that there are degeneracies in the method. Different linear combinations of the base spectra could result in the same continuum. However, in many cases, it is possible to notice the consistence of the spectral synthesis, since it can differentiate spectra with deeper stellar absorptions (associated with old stellar populations) from spectra with flatter stellar absorptions (associated with younger stellar populations) with good precision. It is not possible to determine an uncertainty to those degeneracies, and we are in need of more precise stellar bases and methods to perform this analysis.

\section{Gas kinematics}\label{sec_gaskinematics}

\subsection{General scenario}

In Paper~I, we came out with a general scenario to the main structures of the NGC~613 nucleus in terms of gas emission. There is an optical double emission, separated by a stream of dust, which we called N1 and N2. We believe that the AGN, identified as a variable point-like source (V1), is located between N1 and N2. The eight H\textsc{ii} regions in the circumnuclear ring orbit around this nuclear structure (composed by N1, N2 and V1). As we will see in this work, N1 and N2 are part of the same structure, as we stated in Paper~I. The stream of dust that divides this structure is probably associated with the nuclear spiral that is also located between N1 and N2. 

The centroid of the [O~\textsc{i}]$\lambda$6300 emission, [O~\textsc{i}]$_C$, which is a line typical of regions of partial ionization, is not located at the position of V1 (the AGN). The projected distance between V1 and [O~\textsc{i}]$_C$ is $\sim$~0.24 arcsec. This difference may be due to the ionization cone being partially obscured by dust near the AGN. In Paper~I, [O~\textsc{i}]$_C$ was a reference point to the images superpositions, to compare directly the data from different instruments, since it was the most point-like source in the GMOS data cube (see section 3 of Paper~I, for more details). In the images of this work, we indicate the position of [O~\textsc{i}]$_C$ and V1. 

In this section, we will study the gas dynamics to determine, with better precision, how the structures behave in the nuclear environment of NGC~613. The study is made according to the FOV of each instrument. Since GMOS has the inner central FOV, we are able to study the inner central gas kinematics. In the cases of the SIFS and SINFONI FOVs, we have, besides the central, the circumnuclear gas kinematics in greater detail.

\subsection{Inner central optical gas kinematics (GMOS)}

The GMOS data cube has a FOV that shows the regions that are internal to the circumnuclear ring. Channel maps of emission lines are a good tool to visualize in greater detail the gas kinematics. Fig.~\ref{channelmaphagmos} shows the channel maps of the H$~\alpha$ emission line. As presented in Appendix A of Paper~I, the spectra of the central region of NGC~613 have blended lines, so we have to consider that the first and last channels might be contaminated by the [N~\textsc{ii}]$\lambda$6748 and [N~\textsc{ii}]$\lambda$6784 emission lines, respectively. Between the second and eighth channel maps, we see that there is a gas rotation of the regions at the edges of the FOV, which are the H~\textsc{ii} regions of the circumnuclear ring. We also notice a central emission with a defined orientation between the 8th and 10th channel. This emission is in redshift and probably represents an outflow of gas that comes from V1. In order to represent this outflow, we took the channel with v=~306~km~s$^{-1}$ and the PA of this emission is $\sim$~17$^{\circ}$.

\begin{figure*}
\begin{center}

  \includegraphics[scale=0.3]{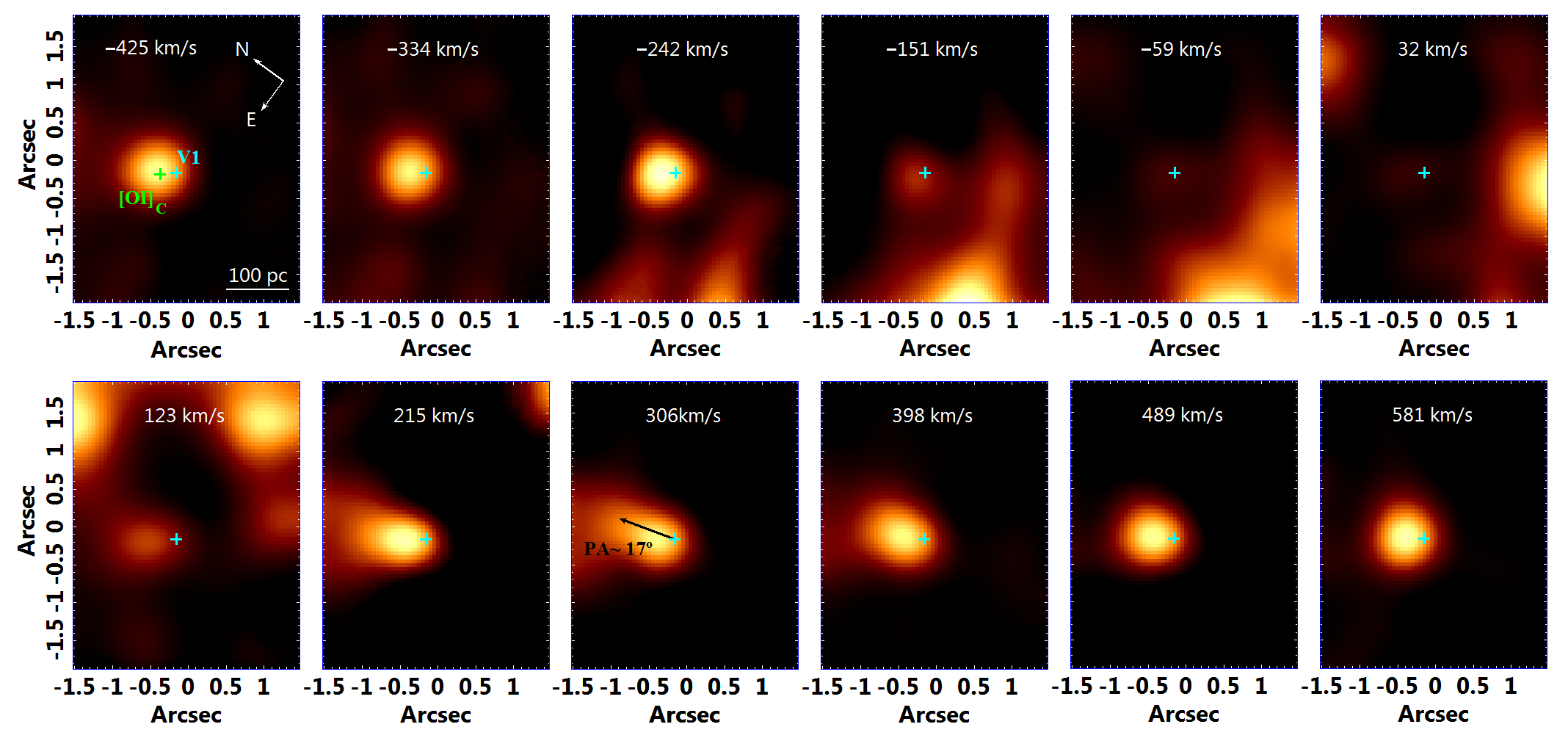}
  \caption{Channel maps of the H~$\alpha$ emission line from the GMOS data cube with 1-\AA\ interval between each channel. The velocities were calculated related to the rest wavelength of the emission line. The positions of [O~\textsc{i}]$_C$ and V1 are indicated by the green and cyan crosses, respectively, and its size represents the 3$\sigma$ uncertainty. The NE orientation of the GMOS data cube is indicated in the first channel, as well as the 100-pc scale. As the H~$\alpha$ emission line seems blended in the central region of NGC~613, the first and last channels might be contaminated with [N~\textsc{ii}]$\lambda$6748 and [N~\textsc{ii}]$\lambda$6784 emission. The channels between 215 and 398~km~s$^{-1}$ show a central emission with a well-defined orientation, which suggests an outflow with redshift velocities. The representative velocity is 306~km~s$^{-1}$ and PA~$\sim$~17$^{\circ}$.   \label{channelmaphagmos}}
  
\end{center}
\end{figure*}

\begin{figure}
\begin{center}

   \includegraphics[scale=0.5]{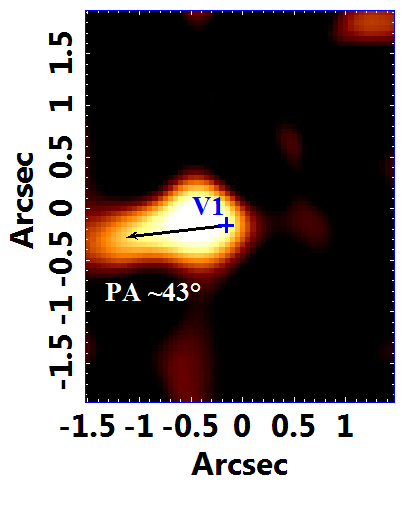} 
  \caption{Image of the [O~\textsc{iii}]$\lambda$5007 emission line with --650~$\lesssim$~v $\lesssim$~--590 km~s$^{-1}$ related to the rest wavelength obtained from the GMOS data cube. This emission has a preferred direction and its is PA $\sim$~43$^{\circ}$. The blue cross represents the position of the AGN (V1) and its size represents the uncertainty of 3$\sigma$. \label{outflowOIII}}
\end{center}
\end{figure}

The channel maps of the [O~\textsc{iii}]$\lambda$5007 emission line of the GMOS data cube are very noisy and hard to interpret due to the low signal-to-noise ratio (S/N) in the blue region of the optical spectra. However, we notice a very clear emission when we observe the channel with  --650~$\lesssim$~v~$\lesssim$~--590 km s$^{-1}$ (see Fig.~\ref{outflowOIII}). This emission has a preferred direction, whose PA~$\sim$~43$^{\circ}$ is not compatible with the one estimated for the outflow observed in the H~$\alpha$ channel maps, neither is the velocity, since it is in blueshift.

\subsection{Near-infrared (NIR) and molecular gas kinematics (SINFONI and ALMA)}

\begin{figure*}
\begin{center}
   \includegraphics[scale=0.5]{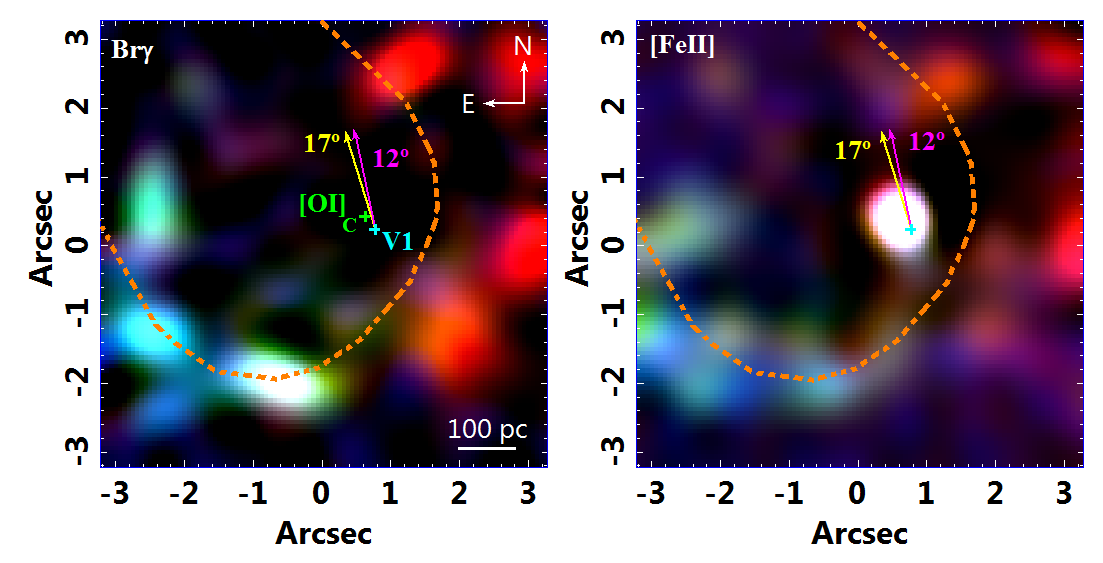}
  \caption{RGB composition of Br~$_{\gamma}$ and [Fe~\textsc{ii}]$\lambda$16436 emission lines from the SINFONI data cube. The velocity intervals for the colours of both images are the same; red: 100~<~v~<~319~km~s$^{-1}$, green:~--49~<~v~<~49~km~s$^{-1}$, and blue: --411~<~v~<~--100~km~s$^{-1}$. The orange contours represent approximately the edges of the ionization cone extracted from the [O~\textsc{iii}]$\lambda$5007 image with --170~km~s$^{-1}$ (see Fig.~\ref{channelmapoiii}), and the magenta and yellow lines indicate, respectively, the PAs of the outflow observed with H~$\alpha$ (PA~$\sim$~17$^{\circ}$) and from the radio jet observed by \citet{hummel} (PA~$\sim$~12$^{\circ}$). The green and cyan crosses represent the positions of [O~\textsc{i}]$_C$ and V1, respectively, and their sizes represent the uncertainty of 3$\sigma$. Note in the Br~$\gamma$ image the indication of the NE orientation and of the 100-pc scale. \label{rgbbrgammafeii}}
\end{center}
\end{figure*}

\begin{figure*}
\begin{center}
   \includegraphics[scale=0.38]{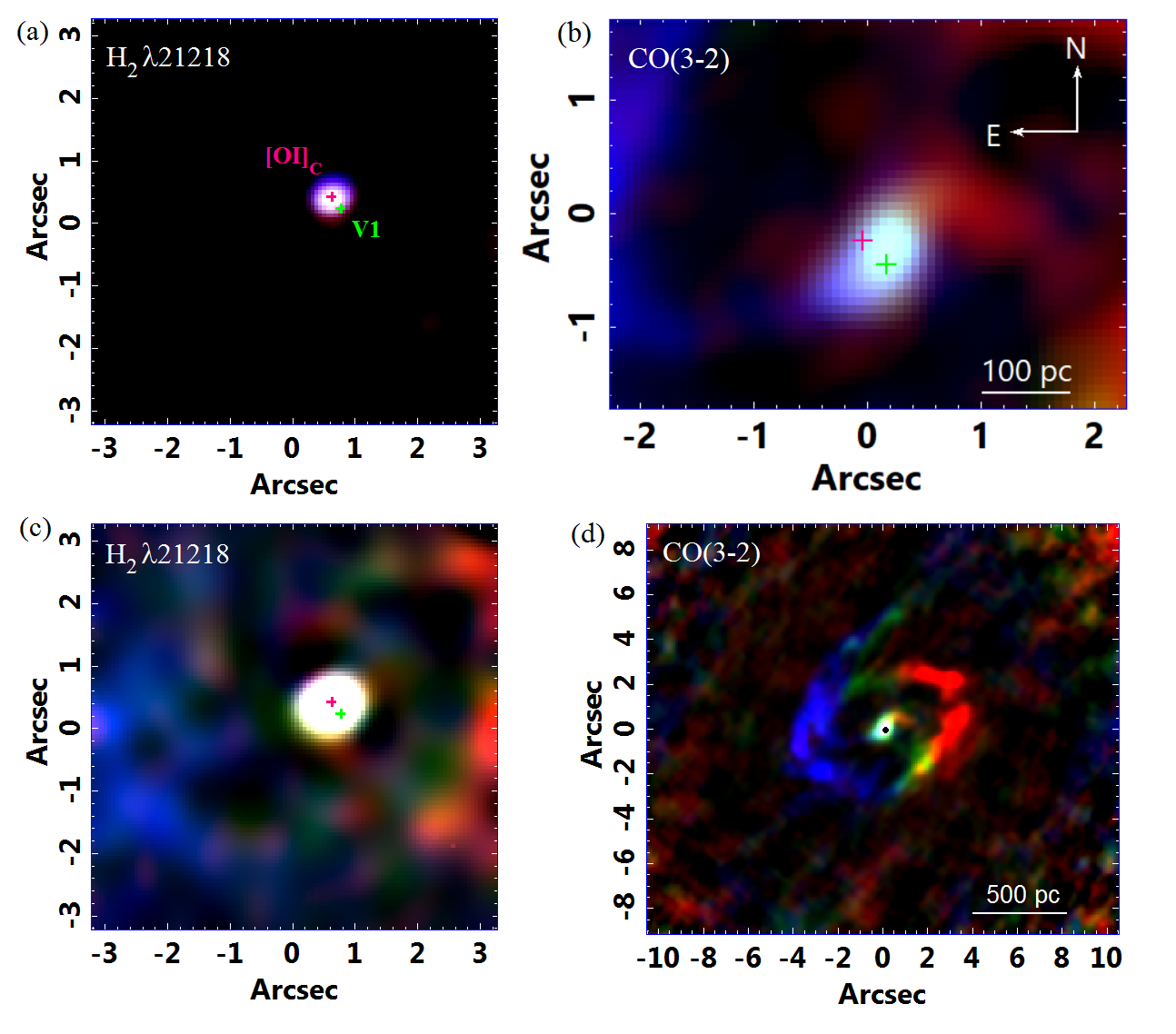}
  \caption{RGB composition of H$_2\lambda$21218 with the LUT (look up table) set up to show only the inner central molecular gas kinematics (panel a) and with the overdone LUT to show the molecular gas kinematics in the circumnuclear ring  (panel c). (Panel b) RGB composition of CO(3-2) emission line observed with ALMA zoomed in the central region and (panel d) with all ALMA FOV. The magenta and green crosses represent the positions of [O~\textsc{i}]$_C$ and V1, respectively, and their sizes represent the 3$\sigma$ uncertainty considering the spaxels of the SINFONI data cube (panels a and c) and ALMA data cube (panels b and d). The black dot on panel (d) represents the position of V1. Its size was enlarged to facilitate the visualization and does not correspond to the uncertainty.\label{rgbsh2}}
\end{center}
\end{figure*}

The SINFONI data cube shows the positions of the H~\textsc{ii} regions with great precision when we look to the Br~$\gamma$ emission line images \citep{boker,falcon613}. Thus, we made an RGB composition of this line to determine the gas kinematics in the circumnuclear ring (see the left-hand panel of Fig.~\ref{rgbbrgammafeii}). We notice that there is a gas rotation of the H~\textsc{ii} regions around the AGN: The regions located to the west present gas emission in redshift and the gas with velocities close to zero seems to be located to the east, together with the gas with emission in blueshift. The [Fe~\textsc{ii}]$\lambda$16436 RGB (right-hand panel of Fig.~ \ref{rgbbrgammafeii}) presents also the same trend; however, the granularity of the image of the H~\textsc{ii} regions is higher than in the Br~$\gamma$ image.

Besides the RGB compositions of [Fe~\textsc{ii}]$\lambda$16436 and Br~$\gamma$, we also created RGBs of the H$_2\lambda$21218 emission line (Fig.~\ref{rgbsh2} panels a and c). In the inner centre of the FOV, we see that the gas with emission in blueshift is located to the north, whereas the gas with emission in redshift is located to the south (Fig.~\ref{rgbsh2}a). That might indicate a rotation of molecular gas around the AGN. When we look at the RGB composition of the ALMA data cube of the inner centre (Fig.~\ref{rgbsh2}b), we see that the nuclear spiral has an opposite pattern in the nucleus: blueshift to the south and redshift to the north (the same trend was detected by \citealt{audibert}). We see also that the nuclear spiral is connected to the bar and its arms have the same kinematics, which suggest that the nuclear spiral is bringing gas and dust to the centre, feeding the AGN. This process has no connection with the molecular gas observed in the H$_2\lambda$21218; therefore, we are presenting two different phenomena. In the ring (panels c and d of Fig.~\ref{rgbsh2}), the molecular gas shows emission (in the circumnuclear ring) in redshift to the west and emission in blueshift to the east, as in the previous RGBs. However the low-velocity gas is located to the west, together with the gas in redshift, the opposite of the RGBs of [Fe~\textsc{ii}]$\lambda$16436 and Br~$\gamma$. In this case, we could not separate completely the images representing higher positive velocities (red colour in the RGBs) from the images corresponding to lower moduli of velocities, due to the insufficient spectral resolution.

\subsection{Circumnuclear optical gas kinematics (SIFS)}

\begin{figure*}
\begin{center}

  \includegraphics[scale=0.29]{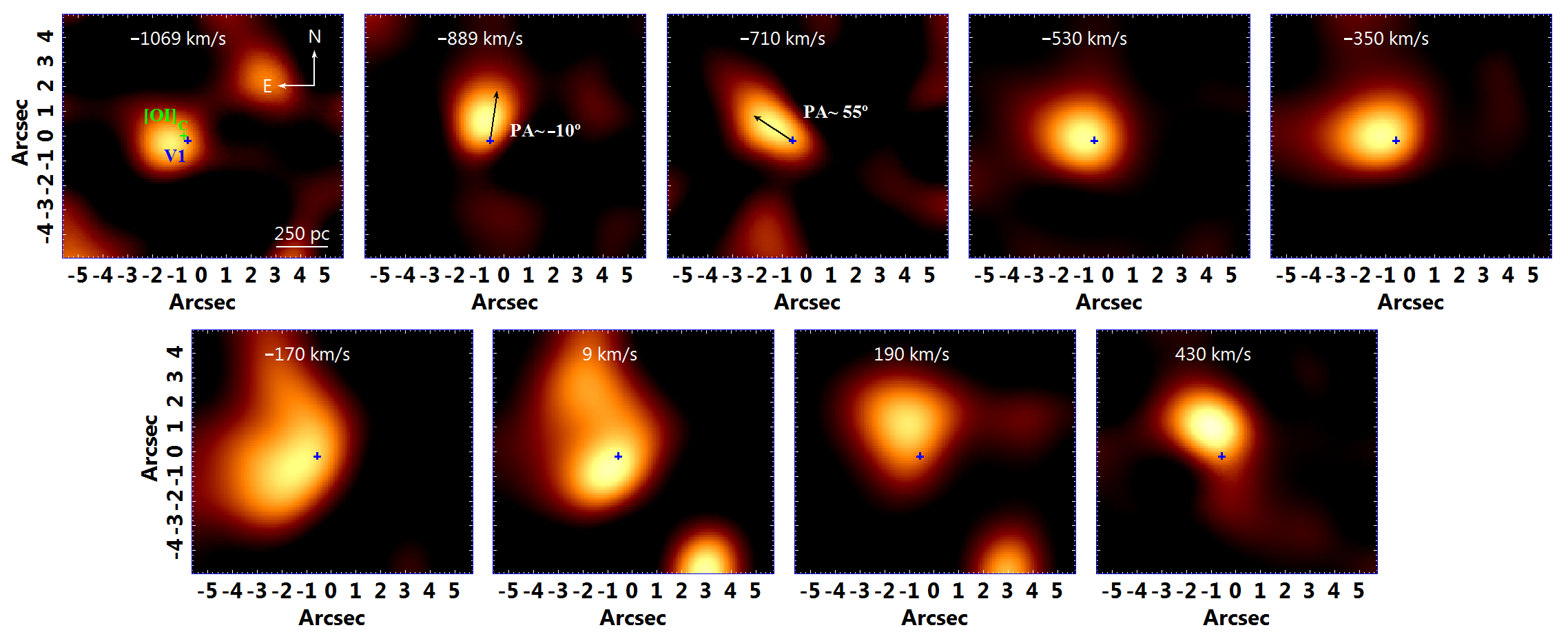}
  \caption{Channel maps of [O~\textsc{iii}]$\lambda$5007 line of the SIFS data cube with 2-\AA\ intervals between each channel. The velocities were determined relative to the rest wavelength of the emission line. The positions of [O~\textsc{i}]$_C$ and V1 are indicated by the green and blue crosses, respectively, and their sizes represent 3$\sigma$ uncertainty. The channels with  v~=~--889 and --710~km~s$^{-1}$ have indications of the PAs of the outflows of gas with PAs~$\sim$~--10\degr and $\sim$~55\degr, respectively. \label{channelmapoiii}}
  
\end{center}
\end{figure*}

One of the most relevant spectral features in the SIFS data cube is the [O~\textsc{iii}]$\lambda$5007 emission line. As we know that there is an AGN (V1), we can associate this emission with an ionization cone that comes from it (see Paper~I). The [O~\textsc{iii}]$\lambda$5007 kinematics was also studied here with the channel maps of Fig.~\ref{channelmapoiii}. The morphology of the ionization cone can be seen in the channels with low velocities (--170 and 9~km~s$^{-1}$). The [O~\textsc{iii}]$\lambda$5007 emission line has a profile with a prominent blue wing, indicating outflows with high negative velocities. The outflow appears clearly in the channel with v~=~--889~km~s$^{-1}$ and PA=~-10\degr. When we compare the channels with velocities --530 and 429~km~s$^{-1}$, we see that there might be a rotation of gas in the inner central region, although we cannot exclude the possibility of other gas outflows. There is no correlation between the [O~\textsc{iii}]$\lambda$5007 emission and the star-forming regions, except that there is a region of a break in the ring that might be associated with the emission of the ionization cone.

\begin{figure*}
\begin{center}

  \includegraphics[scale=0.29]{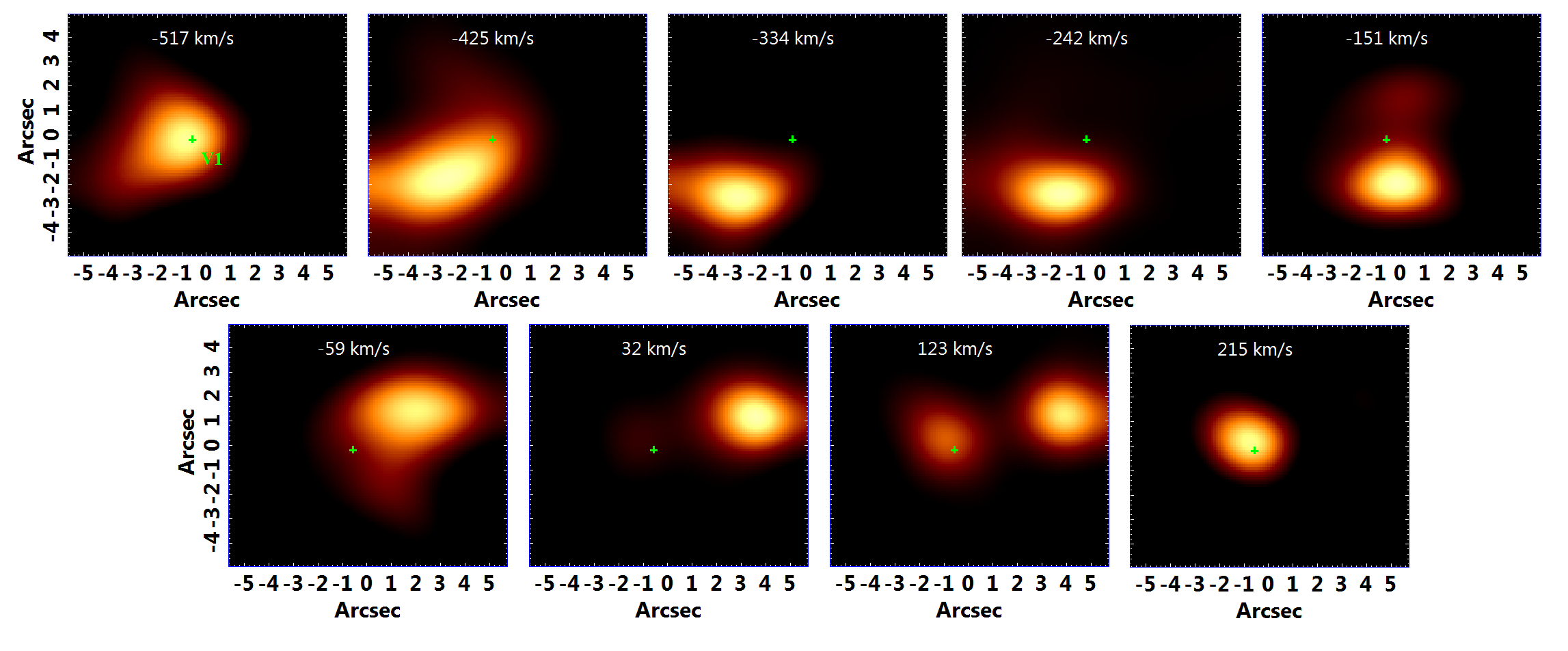}
  \caption{H~$\alpha$ channel maps obtained from SIFS data cube with intervals of 1 \AA\ between each channel. The velocities were estimated relative to the rest wavelength of the emission line. The V1 position is indicated by the green cross, whose size represents the 3$\sigma$ uncertainty. As the H~$\alpha$ line is blended in some regions, it might have some contamination of the [N~\textsc{ii}]$\lambda\lambda$6548,~6584 emission in the first and last channels.\label{channelmaphasifs}}
  
\end{center}
\end{figure*}

We studied the H~$\alpha$ emission-line kinematics of the SIFS data using the channel maps of Fig.~\ref{channelmaphasifs}. As in the GMOS data cube, we have to consider that, in the inner regions of the FOV (mostly in the regions closer to the AGN), the H~$\alpha$ line is blended with the [N~\textsc{ii}]$\lambda\lambda$6548,~6584 lines, so the first and last channels might be contaminated with this emission. When we look at the channel maps of Fig.\ref{channelmaphasifs}, we see that the kinematics of the H~\textsc{ii} regions in the circumnuclear ring (which we cannot spatially separate, due to the low spatial resolution) is consistent with what we see in the SINFONI and ALMA data (see Figs.~\ref{rgbbrgammafeii} and \ref{rgbsh2}). The H~\textsc{ii} regions to the south-east of V1 show emission in blueshift and those to the north-west show emission in redshift.

\section{Stellar archaeology} \label{sec_stellararche}

As said in section \ref{starlight_def}, it is possible to study the stellar archaeology with the spectral synthesis. In this case, we used the \textsc{starlight} software, with a base created from MILES. For more information see section \ref{starlight_def}. 

The spatial resolution of the GMOS data cube is high enough to reveal regions N1 and N2. However, the positions of the H~\textsc{ii} regions cannot be clearly determined from this data cube, as its FOV includes only the inner edges of the circumnuclear ring (3~$\times$~3.8~arcsec$^2$). Thus, the study of the H~\textsc{ii} regions was made with the SIFS data cube, as its FOV is larger (11.4~$\times$~9.9~arcsec$^2$).  

In this section, we present the results of the spectral synthesis applied to the spectra of regions N1 and N2, extracted from the GMOS data cube, and to the spectra of the H~\textsc{ii} regions, extracted from the SIFS data cube.

\subsection{Results from the GMOS data cube}\label{starlight_n1n2}

N1 and N2 spectra were extracted from the GMOS data cube according to Fig.~10 in Paper~I before creating the gas data cube. The diameter of the extraction region was equal to the FWHM of the PSF of the data. 

\begin{figure}
\begin{center}
   \includegraphics[scale=0.4]{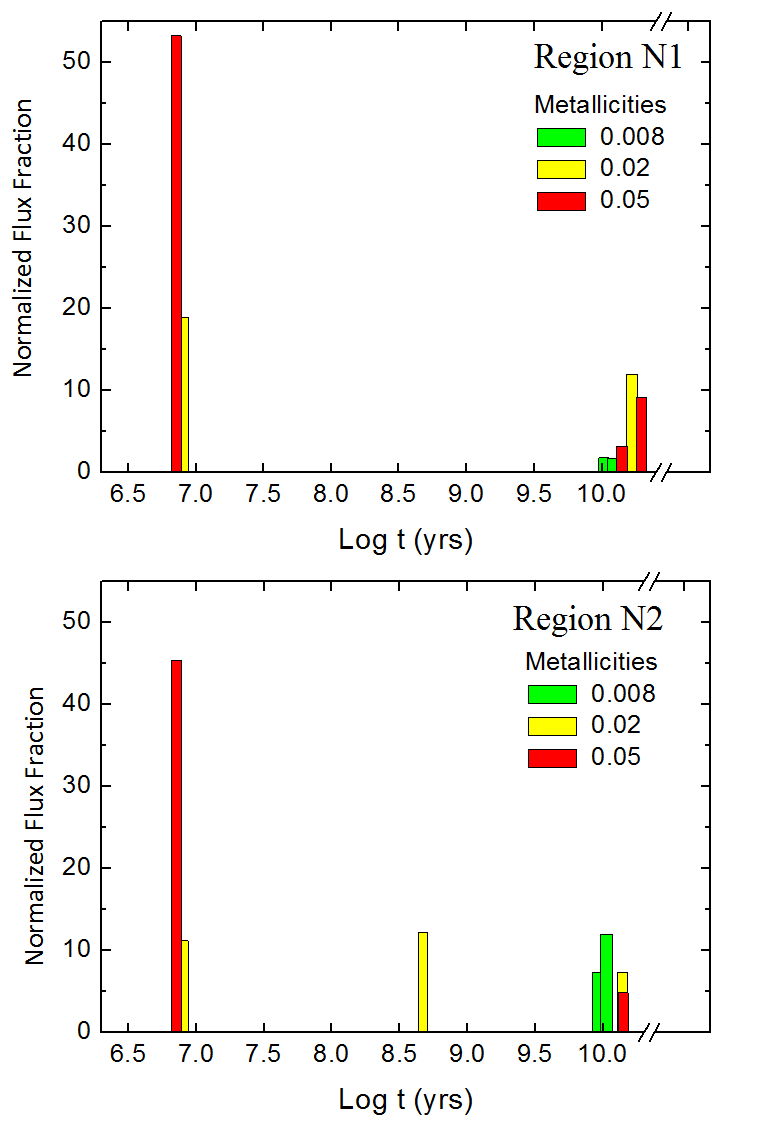}
  \caption{Histograms of the results of the spectral synthesis applied to the N1 and N2 spectra of GMOS data cube. No flux fraction associated with a power law with spectral index equal to 1.5 (representing the featureless continuum of the AGN) was detected (which should be observed in the histogram after the break). \label{histogramasN1N2}}
\end{center}
\end{figure}

From the results of the spectral synthesis, we made histograms of the flux fraction (normalized in each region) of the detected stellar populations. In both regions, the young stellar populations (10$^7$~yr) with high metallicity (0.02 and, mostly, 0.05) are responsible for the highest flux fractions. There are a few differences between regions N1 and N2: the presence of the 10$^{8.5}$-yr stellar populations with 0.02 metallicity and the old stellar populations (10$^{10}$~yr) with intermediate metallicity (0.008) in the N2 region (Fig.~\ref{histogramasN1N2}).

When we compare these data with the SIFS data (see section \ref{starlight_regioesHII}), we see that regions N1 and N2 show flux fractions associated with mainly young stellar populations, while, in the ring, we see a balance between old and young stellar populations.

\begin{figure*}
\begin{center}
   \includegraphics[scale=0.4]{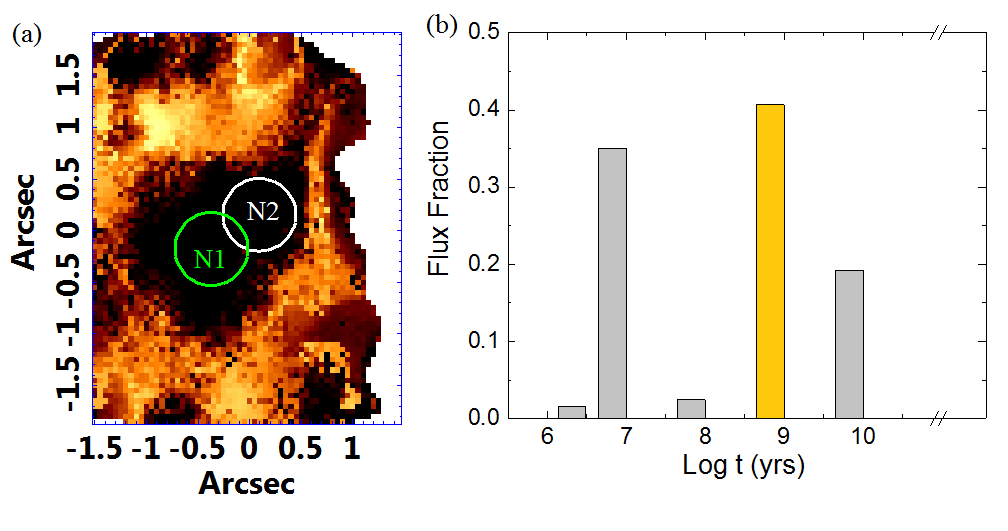}
  \caption{(Panel a) Flux map obtained with the spectral synthesis applied to the whole FOV of the GMOS data cube showing the detected stellar population of 10$^9$~yr. The bright areas of the map show the flux from 10$^9$-yr stellar populations with all metallicity values, but the dominant metallicity is high (0.02 and 0.05). The circles represent regions N1 and N2, and their diameter the FWHM of the PSF of the GMOS data cube. These were the exact extraction areas of the spectra of regions N1 and N2 and, therefore, the regions of the results presented in Fig.~\ref{histogramasN1N2}. The white region of the map was not considered in the results, since it has S/N~$<$~10. (Panel b) Histogram of the results of the spectral synthesis applied to the whole FOV of the GMOS data cube. The flux fraction is not separated by metallicity; it only shows what is the flux fraction associated with each age considered. Only ages with 10$^9$ yr were taken into account in panel~(a), and the others, in grey, are not presented in this work, but in the results of the spectral synthesis of the SIFS data cube in the Appendix \ref{sifs_starlighttotal}. \label{pop109gmos}}
\end{center}
\end{figure*}

Fig.~\ref{pop109gmos}(a) shows the flux map of the populations of 10$^9$~yr detected with the spectral synthesis applied to the entire FOV of the GMOS data cube. The map includes the flux of the 10$^9$-yr stellar populations with all metallicity values, although the dominant metallicity is high (0.02 and 0.05). We notice, by looking at the histograms of Fig.~\ref{histogramasN1N2} and this map, that the concentration of this stellar population is completely circumnuclear and resembles a ring of radius $\sim$~1~arcsec intermediate between the nucleus and the previously known circumnuclear ring (see section~\ref{sec_disc_innerring}). 

The histogram of Fig.~\ref{pop109gmos}(b) shows that the 10$^9$-yr stellar populations are responsible for the largest flux fraction in the inner region of the nucleus of NGC 613, followed by the 10$^7$-yr stellar populations. It also shows how relevant is the inner circumnuclear ring of this intermediate-age stellar populations. A comparison between this histogram and the one in Fig.~\ref{histogramasN1N2} reveals that the 10-Gyr stellar populations are responsible for a larger flux fraction in the circumnuclear region when compared to regions N1 and N2.

\subsection{Results from the SIFS data cube}\label{starlight_regioesHII}

It is already known in the literature that NGC~613 has a circumnuclear star-forming ring \citep{boker,falcon613}. So, from the SIFS data cube, whose FOV covers all the area of the ring (besides the two additional regions detected in this work, see Fig.~10 from Paper~I), we applied a spectral synthesis with the \textsc{starlight} software to the extracted spectra of the H~\textsc{ii} regions named from 1 to 10 (whose extraction areas are shown in Fig.~10 of Paper~I).

\begin{figure*}
\begin{center}
   \includegraphics[scale=0.37]{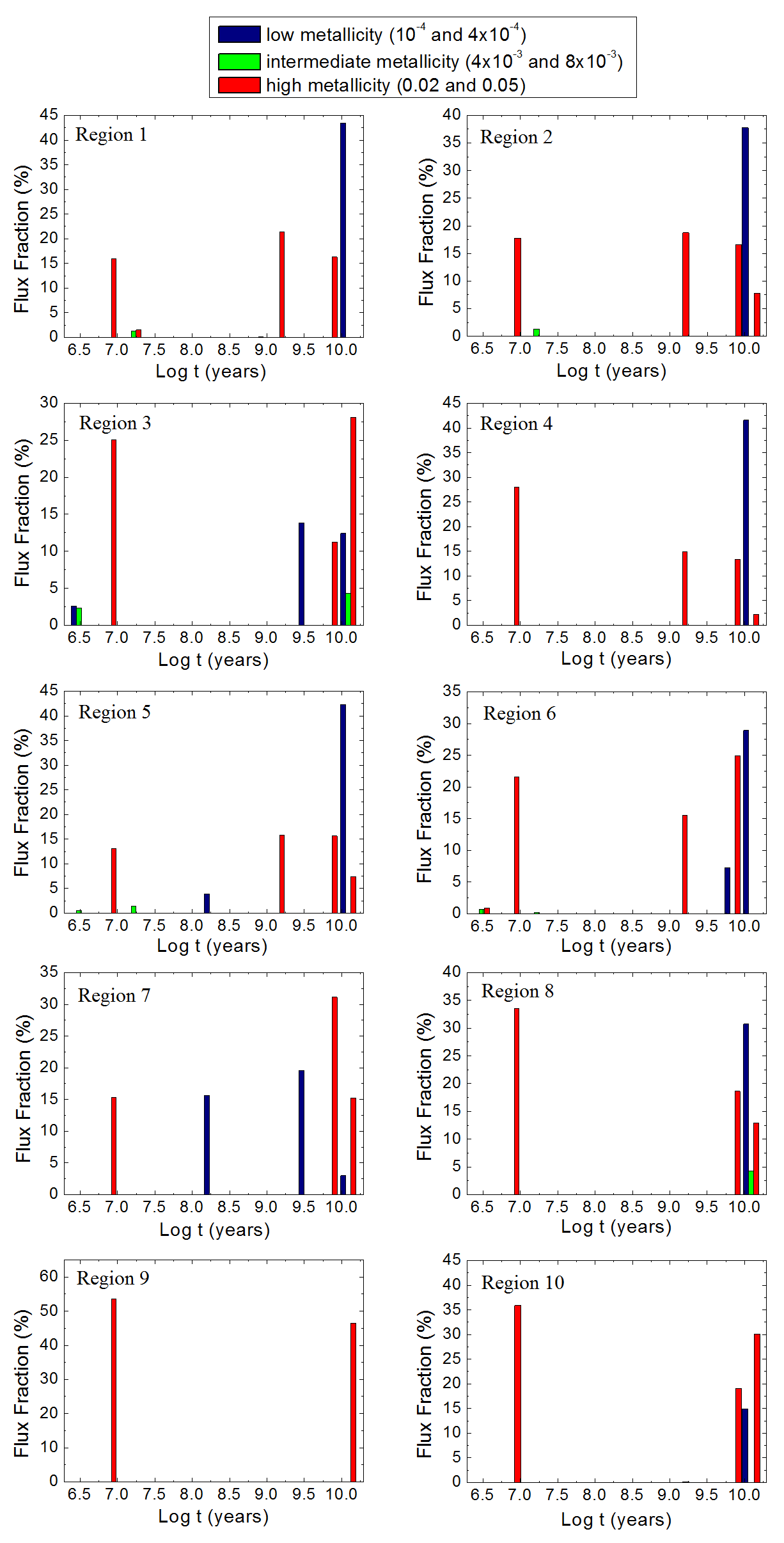}
  \caption{Histograms of the result of the spectral synthesis applied to the extracted spectra of the SIFS data cube of the regions named 1 -- 10 in Fig.~10 of Paper~I. \label{histogramasHII}}
\end{center}
\end{figure*}

We created a histogram from the results of the spectral synthesis applied to the spectrum of each region, as shown in Fig.~\ref{histogramasHII}. We notice that all regions have mainly both young (10$^7$~yr) and old ($\sim$~10$^{10}$~yr) stellar populations, whereas some regions do not show flux fractions associated with intermediate ages. Regions 7, 9, and 10 show most (or all, in the case of region 9) of their flux fractions attributed to high-metallicity stars (0.02 and 0.05). The other regions present, apparently, flux fractions divided between high and low (10$^{-4}$ and $4 \times 10^{-4}$) metallicities.

\section{Stellar kinematics}\label{secstellarkinematics}

It is possible to study the stellar kinematics from the absorption lines using the spectral synthesis. In this case, we used the pPXF method \citep{cappellari}, which is more precise, as the spectral base is convolved with a Gauss--Hermite expansion. In this case, we used the same spectral base used for the stellar archaeology. We then have for results the coefficients of this expansion: $h_1$, the radial velocity, $h_2$, the velocity dispersion and $h_3$, the value that represents the degree of asymmetry of the absorption lines that were fitted. Since this method was applied to data cubes, the results are maps of those coefficients.

The pPXF was applied to the GMOS data cube, in order to study the stellar kinematics of the inner region of NGC~613 nucleus and to the SIFS and SINFONI data cubes, in order to obtain the stellar kinematics of the circumnuclear region (and of the circumnuclear ring).

\subsection{Inner central stellar kinematics obtained from the GMOS data cube}

\begin{figure*}
\begin{center}

  \includegraphics[scale=0.4]{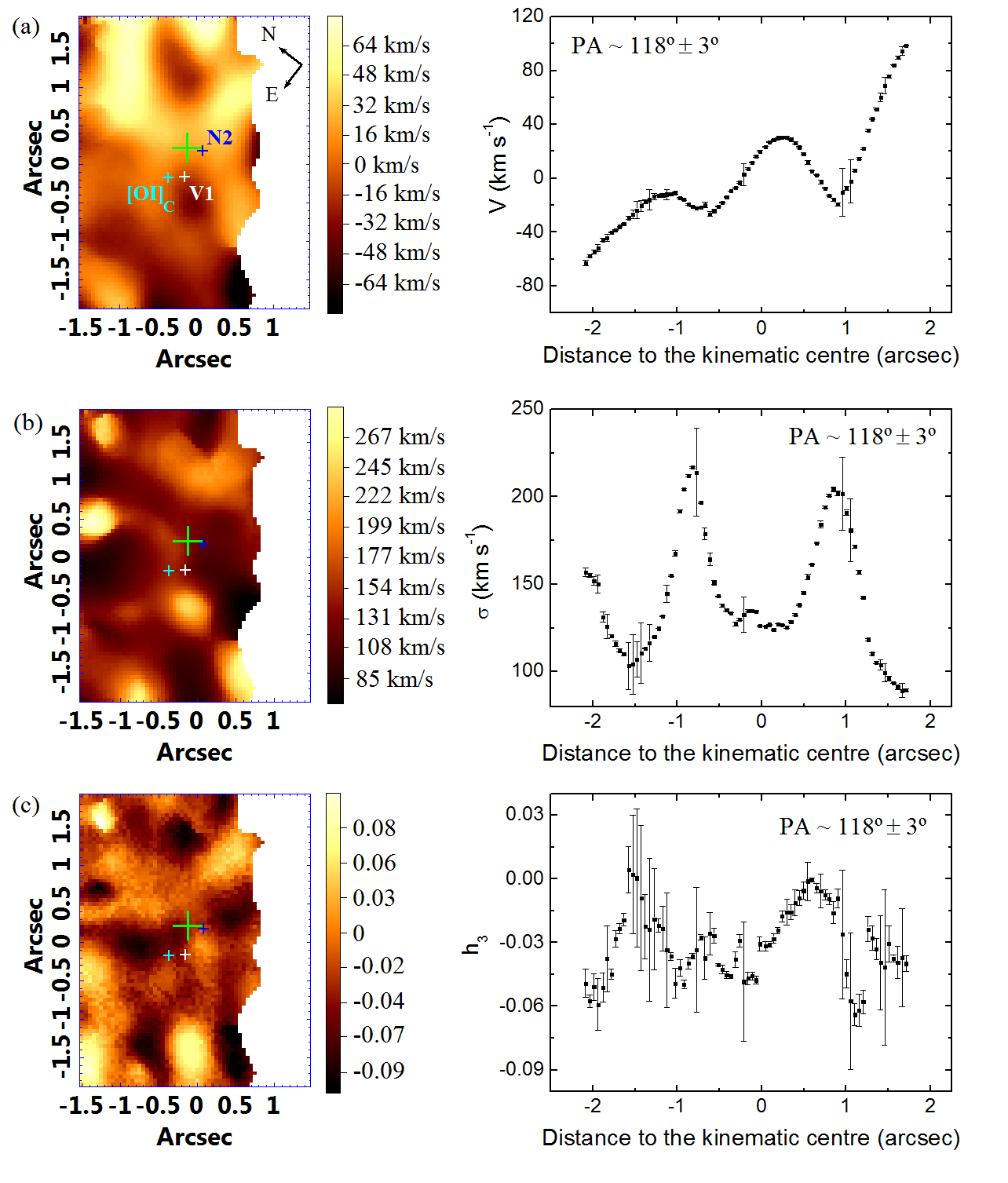}
  \caption{Maps of the coefficients of the Gauss--Hermite expansion of the pPXF applied to the GMOS data cube: (panel a) stellar velocity, (panel b) stellar velocity dispersion and (panel c) h$_3$ coefficient that measures the asymmetry degree of the absorption lines. All maps have their curves extracted along the nodes line, whose PA~$\sim$~--62$^{\circ}$~$\pm$~3$^{\circ}$. The kinematic centre is indicated by the green cross, and its size represents the 1$\sigma$ uncertainty. The cyan, blue, and white crosses indicate the positions of [O~\textsc{i}]$_C$, the centroid of N2, and V1, respectively, and their sizes represents the 3$\sigma$ uncertainty. The areas of the FOV where the S/N~$<$~10 were removed from the map, since their data are not reliable. \label{ppxfgmos}}
  
\end{center}
\end{figure*}

The stellar velocity map of Fig.~\ref{ppxfgmos}(a) is very irregular. One can note a rotation pattern in the inner central 0.5 arcsec. Despite the irregularities, we notice that there is a predominance of negative velocities to the east/south-east and positive velocities to the west/north-west of [O~\textsc{i}]$_C$. The kinematic centre was defined as the point, along the line of nodes, with a velocity value equal to the average between the maximum and minimum velocity values. This average velocity was subtracted from the velocity map, so Fig.~\ref{ppxfgmos}(a) actually shows the radial velocities relative to the kinematic centre, which is between [O~\textsc{i}]$_C$ and N2 and closer to N2. However, considering the uncertainty, the kinematic centre is also compatible with V1, [O~\textsc{i}]$_C$, or even with the N1 centroid, which is not represented in the image.

The stellar velocity dispersion curve has two peaks (Fig.\ref{ppxfgmos}b), one at $\sim$~1 arcsec and the other at $\sim$~--1 arcsec, and the values decrease towards the centre and towards the edges. There are also perturbations in the velocity curve, with values lower than expected, in the case of a Keplerian motion. The positions of such perturbations are coincident with the ones of the velocity dispersion peaks and also with the possible inner circumnuclear ring of 10$^9$~yr stellar populations. The velocity dispersion at the centre of the curve is $\sim$~126~km~s$^{-1}$ and, at the edge, is $\sim$~100~km~s$^{-1}$ (Fig.\ref{ppxfgmos}b). 

In the more peripheral areas of the $h_3$ map, we can see a possible anticorrelation with the velocity map (Fig.~\ref{ppxfgmos}c). The uncertainties in those regions are high, and as the velocity map itself has irregularities, we cannot draw conclusions from this result.

\subsection{Circumnuclear stellar kinematics obtained from the SIFS data cube}

\begin{figure*}
\begin{center}

  \includegraphics[scale=0.38]{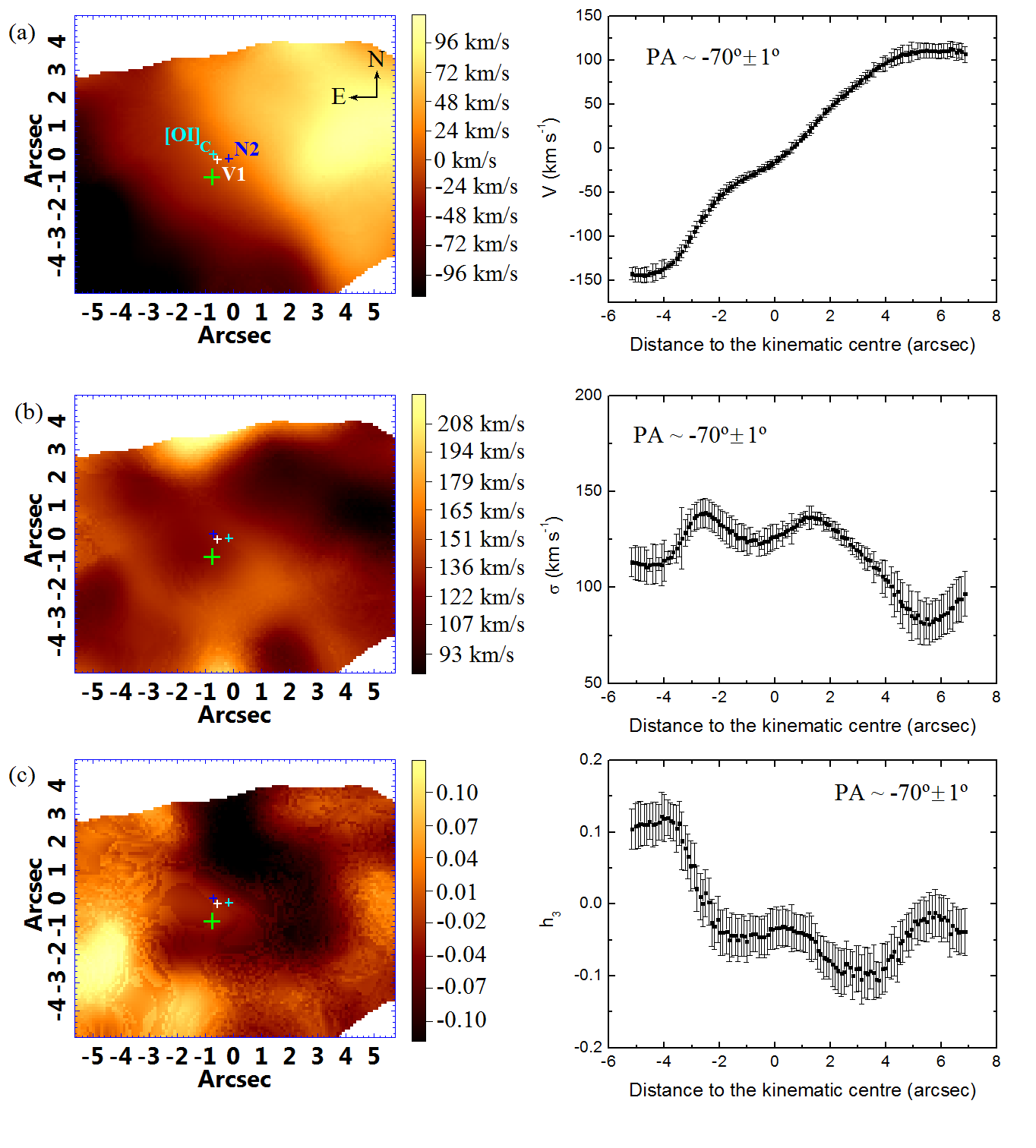}
  \caption{pPXF maps from the SIFS data cube: (panel a) stellar velocity, (panel b) stellar velocity dispersion and (panel c) h$_3$ Gauss--Hermite coefficient, with their respective curves extracted along the nodes line, whose PA~$\sim$~--70$^{\circ}$~$\pm$~1$^{\circ}$. The kinematic centre is indicated by the green cross; the cyan, blue, and white crosses indicate the position of [O~\textsc{i}]$_C$, the centroid of N2, and V1, respectively, and their sizes represent the 3$\sigma$ uncertainty. The areas where the FOV has S/N~$<$~10 were removed from the map. \label{ppxfSIFS}}
  
\end{center}
\end{figure*}

Since it has a larger FOV, the stellar velocity map from the SIFS data cube reveals more clearly the stellar rotation that we could not see in the GMOS results (Fig.~\ref{ppxfSIFS}a). The kinematic centre is not compatible with V1 in this case, being at a distance of 0.66~$\pm$~0.11~arcsec. The stars have negative velocities to the south-east and positive velocities to the north-west. The PA of the line of nodes is --70$^{\circ}$~$\pm$~1$^{\circ}$, compatible at the 3$\sigma$ level with the PA of the line of nodes of the GMOS results, which is --62$^{\circ}$~$\pm$~3$^{\circ}$. 

The velocity dispersion curve of Fig.\ref{ppxfSIFS}(b) shows, as in the GMOS velocity dispersion curve, two peaks at the same position of the perturbations in the velocity curve, although, in the SIFS results, those peaks and perturbations are diluted due to the lower spatial resolution of these data. Those peaks are not located at the same positions of the peaks observed in the GMOS data cube. In fact, their locations are consistent with the circumnuclear ring (not the inner ring as in GMOS). The value of the velocity dispersion at the centre is $\sim$~123~km~s$^{-1}$. 

The $h_3$ map (Fig.~\ref{ppxfSIFS}c) has irregularities but, in general, shows an anticorrelation with the velocity map in the region of the ring, which may indicate that the stars of the ring rotate superposed to a stellar background with velocities close to zero towards the line of sight. 

\subsection{NIR stellar velocity dispersion obtained from SINFONI data cube}

\begin{figure}
\begin{center}

  \includegraphics[scale=0.38]{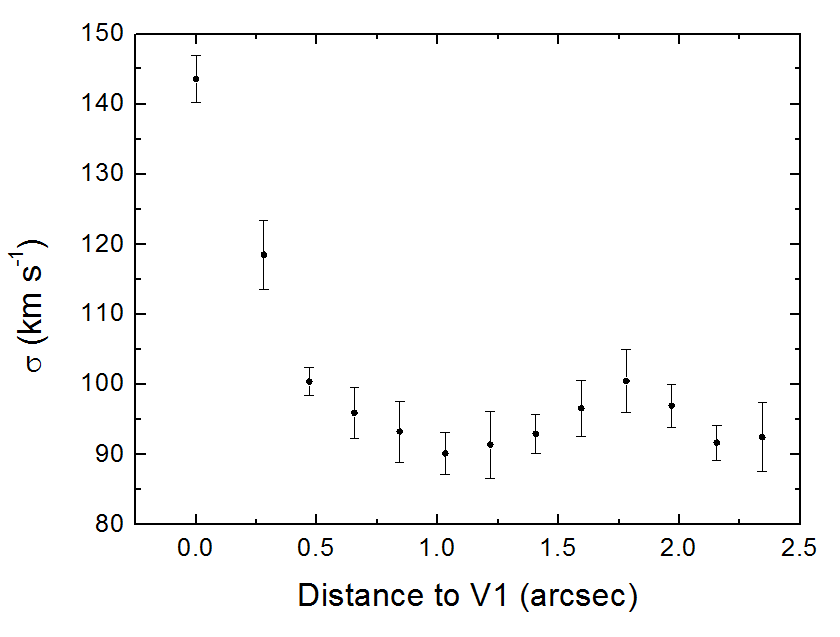}
  \caption{Velocity dispersion curve obtained from the spectra of concentric rings centred at V1 with 3 spaxels of thickness extracted from the SINFONI data cube. \label{dispersaosinfoni}}
  
\end{center}
\end{figure}

Although the SINFONI data cube could have a significant amount of information related to the stellar kinematics, the low S/N prevented the construction of pPXF parameters maps that could be interpreted. In order to obtain, at least, more reliable information about the stellar velocity dispersion, the pPXF method was applied to spectra extracted from concentric rings around V1, with 3 spaxels of thickness. With that, we could minimise the spectral noise and obtain the curve of Fig.~\ref{dispersaosinfoni}.

The uncertainties of the values of the velocity dispersion were obtained by applying a Monte Carlo method to each of the extracted spectra. First, for each spectrum, we subtracted the stellar continuum, using the corresponding pPXF fitting. Then, we created a histogram with the values of the spectrum (after the stellar continuum subtraction) in many spectral intervals, without emission lines. We fitted a Gaussian function to this histogram and created Gaussian distributions of random values with the same width of the Gaussian fitted to the histogram. Those distributions were added to the synthetic stellar spectrum provided by the pPXF, resulting in ``synthetic noisy spectra''. At the end, the pPXF was applied to each one of those ``synthetic noisy spectra'' and the final uncertainty of the stellar velocity dispersion was taken as the standard deviation of all the obtained values.

Unlike what we see in the previous results in the optical, we see here (Fig.~\ref{dispersaosinfoni}) a peak of the stellar velocity dispersion at the centre, with a value of $\sim$~144~km~s$^{-1}$. The stellar velocity dispersion decreases with the radius and reaches a nearly constant value of $\sim~95$~km~s$^{-1}$. At $\sim$~1.8 arcsec from V1, there appears to be a second peak of velocity dispersion, with a value of $\sim~121$~km~s$^{-1}$; however, considering the uncertainties of the values, it is not possible to confirm the existence of this second peak.

In order to obtain a stellar velocity dispersion value representing the central region of NGC~613, we calculated a weighted mean of the values of the map obtained with the pPXF from the SINFONI data cube, taking as weights the integrated fluxes of each spaxel. The value obtained was 92~$\pm$~3~km~s$^{-1}$. The uncertainty was calculated after the creation of two additional maps, one with the sum of the uncertainty map and the other with the subtraction of the uncertainty map. Then, we calculated the weighted means of the dispersion of the obtained maps. The uncertainty was the standard deviation of those values. 

\section{Discussion}\label{sec_discussion}

In this section, we discuss all the data present here, by starting with the scenario of the stellar archaeology and describing the gas and stellar kinematics.

\subsection{Stellar archaeology}

N1 and N2 are two regions with significant stellar emission, so it was convenient to apply the spectral synthesis to their spectra (Fig.~\ref{starlight_n1n2}). The flux fractions of those two regions are due to, essentially, the same stellar populations, without significant differences. We did not detect any region of the data cube with a significant contribution of the featureless continuum of the AGN (represented by a power law with spectral index of 1.5). We actually performed a few tests, applying the spectral synthesis with power laws with different spectral indexes (with values between 1.0 and 1.9); however, we did not find credible evidence for the presence of a featureless continuum, which might indicate that this AGN is so obscured that we cannot detect such an emission. This result is compatible with the conclusions of Paper~I, in which, by analysing the map of D$_{CO}$ created by \citet{falcon613}, we could not see the effect of the featureless continuum that would create a decrease of the values of this map towards the centre (or where the AGN is located). Another hypothesis is related to the fact that the AGN might be a variable source (V1), as we claimed in Paper~I. The absence of a featureless continuum might be due to a low state of activity of the AGN at the epoch of observation. 

The results of the stellar archaeology do not show a continuous star formation in the central region of NGC~613, as only four events of star formation were detected, with most of the H~\textsc{ii} regions showing evidence of just two or three of such events. In the cases of N1 and N2, evidence of two and three star formation events, respectively, was detected. This suggests that the influence of the bar on the star formation in the central region may not be significant, or just might be very recent (10 million years). The absence of a continuous star formation in the nucleus of NGC 613 is not in disagreement with the fact that this galaxy hosts a pseudo-bulge. Actually, as shown by \citet{bre18}, the bulge formation process in a galaxy probably does not follow only two possible routes: one involving a quasi-monolithic gas collapse, which would result in a classical bulge, and the other involving secular processes with continuous star formation, which would form a pseudo-bulge. Instead, the authors conclude that bulges and discs evolve simultaneously in a process that results in a continuous sequence of observed properties, such as stellar age, metallicity, and surface brightness. In addition, it is worth mentioning that studies have shown that the properties of certain pseudo-bulges, specially in isolated galaxies, suggest that most of their mass was formed in an early epoch, without much subsequent star formation \citep{fer14}.

\subsubsection{The intermediate-age inner circumnuclear ring}\label{sec_disc_innerring}

\begin{figure}
\begin{center}

  \includegraphics[scale=0.38]{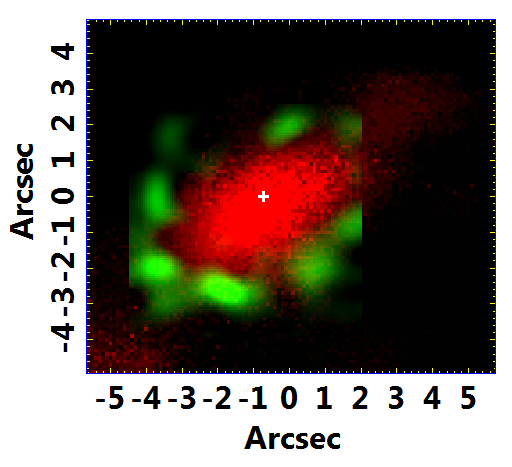}
  \caption{RG composition with the Br~$\gamma$ image showing the circumnuclear ring in green and the image of the flux from the 10$^9$-yr stellar population with all metallicities considered (but predominantly high metallicity –- see Figs.~\ref{histogramasifstotal} and \ref{mapastarlightsifs}c), obtained from the results of the spectral synthesis of the SIFS data cube, in red. One can notice that this stellar population is concentrated inside the region of the ring. The white cross represents the position of V1 and its size represents the uncertainty of 3$\sigma$, considering the size of SIFS spaxels. \label{sifs_anel_109}}
  
\end{center}
\end{figure}

The results of the spectral synthesis applied to the whole FOV of the GMOS data cube show that there is a circumnuclear concentration of the 10$^9$-yr stellar populations (intermediate-ages). By looking only at Fig.~\ref{pop109gmos}, we cannot say if this population is part or not of the circumnuclear ring observed in the NIR, since such a ring is in the FOV edges of the GMOS data cube. We can say that in the inner central regions, this stellar population was not detected, and only in the inner circumnuclear regions. Since we have a wider FOV with the SIFS data cube, we could gather the same flux map of this stellar population (with less inner central resolution because of the differences of the PSFs between GMOS and SIFS data) and compare it directly with the image of Br~$\gamma$ from the SINFONI data cube as shown in Fig.~\ref{sifs_anel_109}. 

We can see that the concentration of this intermediate-ages stellar populations is central, limited to the spatial resolution of the SIFS observations, and it is not part of the circumnuclear ring observed in the Br~$\gamma$ image. It seems that this ring is a barrier of this star formation in the outer regions. The same happens to the 10$^8$-yr stellar populations (Fig. \ref{mapastarlightsifs}b), but the flux fraction of this stellar population is very low and the uncertainty is larger than its value. 

When we compare both results (Figs.~\ref{pop109gmos} and \ref{sifs_anel_109}), we see that the circumnuclear concentration of the 10$^9$-yr stellar population looks like an inner ring, inside the already known circumnuclear ring. So the event that generated this stellar population happened only in the inner region between the nucleus and the circumnuclear ring.

The detection of this inner circumnuclear ring could only be confirmed due to the high spatial resolution of the data. In the SIFS data cube, we can see that the 10$^9$-yr stellar populations are contained inside the circumnuclear ring, but we do not see a ring due to the dilution of the emission caused by the low spatial resolution.

\subsection{Gas kinematics}

The circumnuclear star-forming ring kinematics reveals the same pattern in all the spectral bands and emission lines analysed in this work. The regions located to the east have negative velocities and those to the west have positive velocities (see Figs.~\ref{rgbbrgammafeii} and \ref{rgbsh2}). This result is also compatible with the one obtained by \citet{miyamoto1,miyamoto2} using velocity maps of molecular gas. 

We notice that the nuclear spiral \citep{audibert} has its own kinematics that suggests feeding of the central region. The gas emission of this nuclear spiral to the north-west is in blueshift and that to the south-east in redshift. This same pattern was detected in the H$_2\lambda$21218 emission (see Fig.~\ref{rgbsh2}), which suggests that such a pattern may also be related to the feeding of the central region.

The H~$\alpha$ channel maps of the GMOS data cube reveal the gas kinematics of the inner central region (Fig.\ref{channelmaphagmos}). We notice, at the edges of this FOV, the ring kinematics observed in NIR and ALMA data. Besides that, the outflow we detected, with v~=~306~km~s$^{-1}$, is located at a position compatible with V1 and has a PA~$\sim$~17$^{\circ}$, compatible with the PA of the radio jet observed by \citet{hummel} (see also Fig.~16 of Paper~I). The rotation pattern detected with the H~$\alpha$ channel maps of the SIFS data cube (Fig.~\ref{channelmaphasifs}) is consistent with what we saw in the GMOS, SINFONI, and ALMA data cubes. The outflow of gas, detected in the GMOS data cube, though, cannot be seen here in the SIFS data, due to the lower spatial resolution.

The ionization cone detected in some of the [O~\textsc{iii}]$\lambda$5007 channel maps (v~=~--170 and 9~km~s$^{-1}$) is mostly located to the north-east of V1. The emission detected to the south-west, with velocities higher than --170~km~s$^{-1}$, could be the other side of the ionization cone. The size of the ``break'' in the circumnuclear ring, detected in the Br~$\gamma$, H$_2\lambda$21218, and [Fe~\textsc{ii}]$\lambda$16436 images \citep{falcon613}, north-east of V1, is compatible with the apparent size of the ionization cone. This might indicate that this region is being ionized, along the ionization cone, by the radiation coming from the AGN. However, one of the detected outflows in [O~\textsc{iii}]$\lambda$5007 has a PA that is compatible with the axis of the ionization cone (PA~$\sim$~55$^{\circ}$ and v~=~--710~km~s$^{-1}$, observed with the SIFS data cube, and PA~$\sim$~43$^{\circ}$ and --650~$\lesssim$~v~$\lesssim$~--590~km~s$^{-1}$, observed with the GMOS data cube) and, therefore, within this ''break'' region. The outflow might be pushing the gas out of the ring and possibly preventing star formation in this region or the ionization coming from the AGN is heating the gas at some point where it is not possible for the gas to cool down and form stars. This seems to be just a local impact of the feedback of this AGN in this galaxy, but \citet{daviesbecca} found evidence that the outflowing material from the ionization cone might be interacting with the interstellar medium at larger scales by shocks.

There is a second outflow of gas observed in the [O~\textsc{iii}]$\lambda$5007 channel with velocity of v~=~--889~km~s$^{-1}$, with PA~$\sim$~--10$^{\circ}$. Though this outflow has a PA compatible with the one observed in the H$~\alpha$ channel map and the radio jet, its velocity is very different, so it must be an isolated event or associated with other phenomenon that we did not detect.

According to \citet{hop10}, inflows of gas in a galaxy start at kpc-scales and may be driven by major mergers or even by secular processes (involving the formation of bars). At distances between 1 kpc and 10 pc from the central supermassive black hole (which correspond to the region analysed in this work), the inflow continues, but the gas can have many spatial morphologies, such as spirals, rings, clumps or even secondary bars. The observed properties of the central region of NGC 613 are consistent with this scenario, as the line-emitting regions show different morphologies, such as a nuclear spiral and a circumnuclear ring. As discussed in Section 6.1, this inflow of gas may also result in a gas reservoir for the star formation.

\subsection{Stellar kinematics}

The stellar kinematic pattern detected in this work is essentially the same of the one observed for the gas kinematics along the circumnuclear ring (Figs.~\ref{rgbbrgammafeii} and \ref{rgbsh2}): blueshift to the south-east and redshift to the north-west of V1 in the velocity maps obtained from the pPXF of the SIFS data cube (the only cube that allowed the determination of the kinematics along the entire ring extension -- see Fig.~\ref{ppxfSIFS}a). However, in order to establish a more precise comparison between the stellar and gas kinematics along the circumnuclear ring, we made a radial velocity map of the gas by fitting a Gaussian function to the H~$\alpha$ emission line. We verified that the PA of the line of nodes and also the curve extracted along the line of nodes for both the stellar and gas kinematics are consistent. This indicates that the stars and the gas along the circumnuclear ring share the same kinematics. 

The velocity maps obtained with the GMOS and SIFS data cubes (Figs.~\ref{ppxfgmos}a and \ref{ppxfSIFS}a, respectively) show evidence of perturbations (similar to "depressions"), which are probably related to the two peaks observed in the velocity dispersion curves (Figs.~\ref{ppxfgmos}b and \ref{ppxfSIFS}b). As said previously, the positions of these peaks coincide with the positions of the inner circumnuclear ring (in the GMOS results) and of the circumnuclear ring (in the SIFS results). A possible explanation for this behaviour is that the observed rings have lower radial velocities when compared to the bulge stars. Therefore, the superposition of the spectra of the stellar populations in the bulge and in the rings results in the increase of the velocity dispersion values.

The stellar velocity dispersion maximum determined from the curve obtained from the SINFONI data cube ($\sim~144$~km~s$^{-1}$) is compatible with the ones obtained by \citet{schechter} and \citet{batcheldor} (using the maximum penalized likelihood method), while our average value of the stellar velocity dispersion taken from the SINFONI data cube (92~$\pm$~3~km~s$^{-1}$) is compatible with the value obtained by \citet{batcheldor} using the cross-correlation method. By observing the velocity dispersion curve of SINFONI, we can see that there is one central peak, followed by a higher value at $\sim$0.3 arcsec and a series of values that, within the uncertainties, are constant. We do not see clearly the perturbations mentioned before and it is probably due to the fact that those stellar populations have some different effects when we look at the NIR spectrum.

Comparisons between active and non-active galaxies suggest that active galaxies show, in their central regions, lower values of stellar velocity dispersion than non-active galaxies. \citet{hicks13} found in their sample of active galaxies that the values of the stellar velocity dispersion did not exceed 100 km s$^{-1}$ (see fig. 17 from \citealt{hicks13}). This might be due to the presence of young stars that inherited the cold kinematics of their parent clouds, which probably constitute a gas reservoir for the star formation and also for the accretion onto the central supermassive black hole. In this case, the central region of NGC 613 shows stellar velocity dispersion values more compatible with the ones of active galaxies (in the \citealt{hicks13} sample) and also the presence of young stellar populations. In addition, the presence of the circumnuclear rings and the evidence of inflows of gas help to corroborate this idea.

The portion of the ring with negative velocities has positive values of $h_3$, while the positive velocity regions of the ring have negative values of $h_3$. This anticorrelation of velocity and values of $h_3$ could indicate that the stars of the ring rotate around the nucleus superposed to an stellar background with stars with velocities close to zero. Such anticorrelation is not seen to the north-west, which might indicate deviations of a simple stellar rotation around the nucleus.

By using the stellar velocity map obtained from the SIFS data cube, we could estimate a representative velocity for the stars of the circumnuclear ring, v~$\sim$~100~km~s$^{-1}$, rotating around the nucleus. Considering this velocity and the average values of the ring radius and inclination taken from Table~4 of Paper~I ($\bar{R}$~=~249~pc and $\bar{i}$~=~57$^{\circ}$), we conclude that the time necessary for a group of stars to complete a turn around the ring is $\sim$~8~$\times~10^6$~yr. The temporal resolution of the spectral synthesis method is lower than this value. The scenario ''pearls on string''  \citep{BOKERIAU,boker,falcon613} predicts that there are two gas concentrations at opposite points of the ring that form H~\textsc{ii} regions. Such H~\textsc{ii} regions rotate in the ring and this rotation generates an age gradient along the ring. Considering the estimated time to a full turn of a group of stars along the ring and the temporal resolution of the spectral synthesis, we conclude that more precise methods are necessary in order to determine stellar ages in this case, to confirm this scenario.

\section{Conclusions}\label{sec_conclusion}

In this work, we continue the analysis of the NGC~613 nucleus started in Paper I, focusing on the kinematics and on the stellar archaeology. The main findings are as follows:

(i) From channel maps of the H~$\alpha$ emission line obtained from the GMOS-IFU data cube, we detected an outflow in redshift (with velocity $\sim$~300~km~s$^{-1}$), with the PA ($\sim$~17$^{\circ}$) compatible with the PA of the radio jet (PA~$\sim$~12$^{\circ}$).

(ii) We also detected two outflows of gas also from channel maps of [O~\textsc{iii}]$\lambda$5007. One of them is probably associated with the ionization cone, having the PA~$\sim$~55$^{\circ}$ and v$\sim$~--710~km~s$^{-1}$ (SIFS) and PA~$\sim$~43$^{\circ}$ and v$\sim$~--620~km~s$^{-1}$ (GMOS), respectively, in blueshift. This outflow might be responsible for the ``break" of the circumnuclear ring, observed in the Br$~\gamma$ image, near Region 8, expelling the gas of the circumnuclear ring and preventing star formation. The other outflow has PA~$\sim$~--10$^{\circ}$ with v$\sim$~--890~km~s$^{-1}$, also in blueshift.

(iii) The spectral synthesis results confirm that there is no featureless continuum emission detected in the optical data cubes. This agrees with the absence of decreasing values of the D$_{CO}$ in the NIR towards the centre, as discussed in Paper I.

(iv) The spectral synthesis of the optical data shows that there is no significant difference between the stellar populations of N1 and N2, formed mainly of high-metallicity stars (0.02 and 0.05) with ages of $\sim$~10~Myr and $\sim$~10~Gyr. Such result supports the hypothesis that N1 and N2 are part of the same central structure, apparently separated by a dust lane (probably associated with the nuclear spiral). Another result that reinforces this idea is that there was no measurable difference in the results of the stellar velocity between N1 and N2. This suggests that any significant kinematics, if there is any, might be occurring perpendicular to the line of sight.

(v) The spectral synthesis of optical data shows that the H~\textsc{ii} regions that are part of the circumnuclear ring have young stellar populations with similar ages. We do not have enough temporal resolution to determine age differences between them.

(vi) The star formation history of the NGC~613 centre is not continuous. In all regions of the circumnuclear ring (except Region 8), besides the 10$^{10}$-yr star formation (old stellar populations), another event of star formation happened between $\sim$~1 and $\sim$~3~Gyr (intermediate-age stellar populations), and after that, there was an interval of $\sim$~9~$\times$~10$^{9}$ yr  without star formation (except in Region 7). Outside the ring (regions 9 and 10), this interval of no star formation is even bigger: $\sim$~10$^{10}$~yr.

(vii) A concentration of the 10$^9$-yr (intermediate-age) high-metallicity stellar population was detected in the inner central region. This resembles a ring between the already known circumnuclear ring and the nucleus. This inner circumnuclear ring may be responsible for the observed inner two peaks (perturbations) of stellar velocity and stellar velocity dispersion in the GMOS stellar kinematics results.

(viii) A stellar rotation of the H~\textsc{ii} regions around the centre was detected from the kinematic maps obtained with the pPXF method. This rotation is compatible with what was already observed in Br~$\gamma$ and molecular gas kinematics. 

(ix) The stellar velocity dispersion values obtained from the SINFONI data cube show a central peak with $\sim$~144~km~s$^{-1}$. The average central value for the velocity dispersion, weighted by flux, calculated in all FOV of SINFONI data cube is 92~$\pm$~3~km~s$^{-1}$.

\section*{Acknowledgements}

This work is based on observations obtained at the Gemini Observatory (processed using the Gemini \textsc{iraf} package), which is operated by the Association of Universities for Research in Astronomy, Inc., under a cooperative agreement with the National Science Foundation (NSF) on behalf of the Gemini partnership: the NSF (United States), the National Research Council (Canada), Comisi\'on Nacional de Investigaci\'on Cient\'ifica y Tecnol\'ogica (Chile), the Australian Research Council (Australia), Minist\'erio da Ci\^encia, Tecnologia e Inova\c{c}\~ao (Brazil), and Ministerio de Ciencia, Tecnolog\'ia e Innovaci\'on Productiva (Argentina). This research has also made use of the NASA/IPAC Extragalactic Database (NED), which is operated by the Jet Propulsion Laboratory, California Institute of Technology, under contract with the National Aeronautics and Space Administration. This paper also makes use of the following Atacama Large Millimeter/submillimeter Array (ALMA) data: ADS/JAO.ALMA 2015.1.00404.S. ALMA is a partnership of European Southern Observatory (ESO; representing its member states), NSF (USA), and the National Institutes of Natural Sciences (NINS; Japan), together with National Research Council (NRC; Canada), Ministry of Science and Technology (MOST) and Academia Sinica Institute of Astronomy and Astrophysics (ASIAA; Taiwan), and Korea Astronomy and Space Science Institute (KASI; Republic of Korea), in cooperation with the Republic of Chile. The Joint ALMA Observatory is operated by ESO, Associated Universities, Inc. and the National Radio Astronomy Observatory (AUI/NRAO), and the National Astronomical Observatory of Japan (NAOJ). We thank CNPq (Conselho Nacional de Desenvolvimento Cient\'ifico e Tecnol\'ogico), under grant 141766/2016-6, and FAPESP (Funda\c{c}\~ao de Amparo \`a Pesquisa do Estado de S\~ao Paulo), under grant 2011/51680-6, for supporting this work.

%%%%%%%%%%%%%%%%%%%%%%%%%%%%%%%%%%%%%%%%%%%%%%%%%%



\bibliographystyle{mnras}
\bibliography{references} % if your bibtex file is called example.bib

\begin{thebibliography}{}
\makeatletter
\relax
\def\mn@urlcharsother{\let\do\@makeother \do\$\do\&\do\#\do\^\do\_\do\%\do\~}
\def\mn@doi{\begingroup\mn@urlcharsother \@ifnextchar [ {\mn@doi@}
  {\mn@doi@[]}}
\def\mn@doi@[#1]#2{\def\@tempa{#1}\ifx\@tempa\@empty \href
  {http://dx.doi.org/#2} {doi:#2}\else \href {http://dx.doi.org/#2} {#1}\fi
  \endgroup}
\def\mn@eprint#1#2{\mn@eprint@#1:#2::\@nil}
\def\mn@eprint@arXiv#1{\href {http://arxiv.org/abs/#1} {{\tt arXiv:#1}}}
\def\mn@eprint@dblp#1{\href {http://dblp.uni-trier.de/rec/bibtex/#1.xml}
  {dblp:#1}}
\def\mn@eprint@#1:#2:#3:#4\@nil{\def\@tempa {#1}\def\@tempb {#2}\def\@tempc
  {#3}\ifx \@tempc \@empty \let \@tempc \@tempb \let \@tempb \@tempa \fi \ifx
  \@tempb \@empty \def\@tempb {arXiv}\fi \@ifundefined
  {mn@eprint@\@tempb}{\@tempb:\@tempc}{\expandafter \expandafter \csname
  mn@eprint@\@tempb\endcsname \expandafter{\@tempc}}}

\bibitem[\protect\citeauthoryear{{Alloin} \& {Kunth}}{{Alloin} \&
  {Kunth}}{1979}]{alloin}
{Alloin} D.,  {Kunth} D.,  1979, \aap, \href
  {https://ui.adsabs.harvard.edu/#abs/1979A&A....71..335A} {71, 335}

\bibitem[\protect\citeauthoryear{{Asmus}, {Gandhi}, {H{\"o}nig}, {Smette}  \&
  {Duschl}}{{Asmus} et~al.}{2015}]{ASMUS2015}
{Asmus} D.,  {Gandhi} P.,  {H{\"o}nig} S.~F.,  {Smette} A.,   {Duschl} W.~J.,
  2015, \mn@doi [\mnras] {10.1093/mnras/stv1950}, \href
  {https://ui.adsabs.harvard.edu/\#abs/2015MNRAS.454..766A} {454, 766}

\bibitem[\protect\citeauthoryear{{Audibert} et~al.,}{{Audibert}
  et~al.}{2019}]{audibert}
{Audibert} A.,  et~al., 2019, \mn@doi [\aap] {10.1051/0004-6361/201935845},
  \href {https://ui.adsabs.harvard.edu/abs/2019A&A...632A..33A} {632, A33}

\bibitem[\protect\citeauthoryear{{Batcheldor} et~al.,}{{Batcheldor}
  et~al.}{2005}]{batcheldor}
{Batcheldor} D.,  et~al., 2005, \mn@doi [The Astrophysical Journal Supplement
  Series] {10.1086/431483}, \href
  {https://ui.adsabs.harvard.edu/\#abs/2005ApJS..160...76B} {160, 76}

\bibitem[\protect\citeauthoryear{{B{\"o}ker}, {Falcon-Barroso}, {Knapen},
  {Schinnerer}, {Allard}  \& {Ryder}}{{B{\"o}ker} et~al.}{2007}]{BOKERIAU}
{B{\"o}ker} T.,  {Falcon-Barroso} J.,  {Knapen} J.~H.,  {Schinnerer} E.,
  {Allard} E.,   {Ryder} S.,  2007, in {Vazdekis} A.,  {Peletier} R.,  eds,
  IAU Symposium Vol. 241, Stellar Populations as Building Blocks of Galaxies.
  pp 497--498, \mn@doi{10.1017/S1743921307008873}

\bibitem[\protect\citeauthoryear{{B{\"o}ker}, {Falc{\'o}n-Barroso},
  {Schinnerer}, {Knapen}  \& {Ryder}}{{B{\"o}ker} et~al.}{2008}]{boker}
{B{\"o}ker} T.,  {Falc{\'o}n-Barroso} J.,  {Schinnerer} E.,  {Knapen} J.~H.,
  {Ryder} S.,  2008, \mn@doi [\aj] {10.1088/0004-6256/135/2/479}, \href
  {http://adsabs.harvard.edu/abs/2008AJ....135..479B} {135, 479}

\bibitem[\protect\citeauthoryear{{Breda} \& {Papaderos}}{{Breda} \&
  {Papaderos}}{2018}]{bre18}
{Breda} I.,  {Papaderos} P.,  2018, \mn@doi [\aap]
  {10.1051/0004-6361/201731705}, \href
  {https://ui.adsabs.harvard.edu/abs/2018A&A...614A..48B} {614, A48}

\bibitem[\protect\citeauthoryear{{Cappellari} \& {Emsellem}}{{Cappellari} \&
  {Emsellem}}{2004}]{cappellari}
{Cappellari} M.,  {Emsellem} E.,  2004, \mn@doi [\pasp] {10.1086/381875}, \href
  {http://adsabs.harvard.edu/abs/2004PASP..116..138C} {116, 138}

\bibitem[\protect\citeauthoryear{{Castangia}, {Panessa}, {Henkel}, {Kadler}  \&
  {Tarchi}}{{Castangia} et~al.}{2013}]{castangia2013}
{Castangia} P.,  {Panessa} F.,  {Henkel} C.,  {Kadler} M.,   {Tarchi} A.,
  2013, \mn@doi [\mnras] {10.1093/mnras/stt1824}, \href
  {https://ui.adsabs.harvard.edu/\#abs/2013MNRAS.436.3388C} {436, 3388}

\bibitem[\protect\citeauthoryear{{Cid Fernandes}, {Mateus}, {Sodr{\'e}},
  {Stasi{\'n}ska}  \& {Gomes}}{{Cid Fernandes} et~al.}{2005}]{starlight}
{Cid Fernandes} R.,  {Mateus} A.,  {Sodr{\'e}} L.,  {Stasi{\'n}ska} G.,
  {Gomes} J.~M.,  2005, \mn@doi [\mnras] {10.1111/j.1365-2966.2005.08752.x},
  \href {http://adsabs.harvard.edu/abs/2005MNRAS.358..363C} {358, 363}

\bibitem[\protect\citeauthoryear{{Combes} et~al.,}{{Combes}
  et~al.}{2019}]{combes613}
{Combes} F.,  et~al., 2019, \mn@doi [\aap] {10.1051/0004-6361/201834560}, \href
  {https://ui.adsabs.harvard.edu/abs/2019A&A...623A..79C} {623, A79}

\bibitem[\protect\citeauthoryear{{Davies} et~al.,}{{Davies}
  et~al.}{2017}]{daviesbecca}
{Davies} R.~L.,  et~al., 2017, \mn@doi [\mnras] {10.1093/mnras/stx1559}, \href
  {http://adsabs.harvard.edu/abs/2017MNRAS.470.4974D} {470, 4974}

\bibitem[\protect\citeauthoryear{{Falc{\'o}n-Barroso}, {B{\"o}ker},
  {Schinnerer}, {Knapen}  \& {Ryder}}{{Falc{\'o}n-Barroso}
  et~al.}{2008}]{falconIAU}
{Falc{\'o}n-Barroso} J.,  {B{\"o}ker} T.,  {Schinnerer} E.,  {Knapen} J.~H.,
  {Ryder} S.,  2008, in {Bureau} M.,  {Athanassoula} E.,   {Barbuy} B.,  eds,
  IAU Symposium Vol. 245, Formation and Evolution of Galaxy Bulges. pp 177--180
  (\mn@eprint {arXiv} {0709.0353}), \mn@doi{10.1017/S1743921308017584}

\bibitem[\protect\citeauthoryear{{Falc{\'o}n-Barroso}, {Ramos Almeida},
  {B{\"o}ker}, {Schinnerer}, {Knapen}, {Lan{\c c}on}  \&
  {Ryder}}{{Falc{\'o}n-Barroso} et~al.}{2014}]{falcon613}
{Falc{\'o}n-Barroso} J.,  {Ramos Almeida} C.,  {B{\"o}ker} T.,  {Schinnerer}
  E.,  {Knapen} J.~H.,  {Lan{\c c}on} A.,   {Ryder} S.,  2014, \mn@doi [\mnras]
  {10.1093/mnras/stt2189}, \href
  {http://adsabs.harvard.edu/abs/2014MNRAS.438..329F} {438, 329}

\bibitem[\protect\citeauthoryear{{Fern{\'a}ndez Lorenzo}
  et~al.,}{{Fern{\'a}ndez Lorenzo} et~al.}{2014}]{fer14}
{Fern{\'a}ndez Lorenzo} M.,  et~al., 2014, \mn@doi [\apjl]
  {10.1088/2041-8205/788/2/L39}, \href
  {https://ui.adsabs.harvard.edu/abs/2014ApJ...788L..39F} {788, L39}

\bibitem[\protect\citeauthoryear{{Gonzalez} \& {Woods}}{{Gonzalez} \&
  {Woods}}{2002}]{gwoods}
{Gonzalez} R.~C.,  {Woods} R.~E.,  2002, {Digital image processing (2nd ed.,
  Upper Saddle River, NJ: Prentice Hall)}

\bibitem[\protect\citeauthoryear{{Hicks}, {Davies}, {Maciejewski}, {Emsellem},
  {Malkan}, {Dumas}, {M{\"u}ller-S{\'a}nchez}  \& {Rivers}}{{Hicks}
  et~al.}{2013}]{hicks13}
{Hicks} E.~K.~S.,  {Davies} R.~I.,  {Maciejewski} W.,  {Emsellem} E.,  {Malkan}
  M.~A.,  {Dumas} G.,  {M{\"u}ller-S{\'a}nchez} F.,   {Rivers} A.,  2013,
  \mn@doi [\apj] {10.1088/0004-637X/768/2/107}, \href
  {https://ui.adsabs.harvard.edu/abs/2013ApJ...768..107H} {768, 107}

\bibitem[\protect\citeauthoryear{{Hopkins} \& {Quataert}}{{Hopkins} \&
  {Quataert}}{2010}]{hop10}
{Hopkins} P.~F.,  {Quataert} E.,  2010, \mn@doi [\mnras]
  {10.1111/j.1365-2966.2010.17064.x}, \href
  {https://ui.adsabs.harvard.edu/abs/2010MNRAS.407.1529H} {407, 1529}

\bibitem[\protect\citeauthoryear{{Hummel} \& {Jorsater}}{{Hummel} \&
  {Jorsater}}{1992}]{hummel}
{Hummel} E.,  {Jorsater} S.,  1992, \aap, \href
  {http://adsabs.harvard.edu/abs/1992A%26A...261...85H} {261, 85}

\bibitem[\protect\citeauthoryear{{Hummel}, {Jorsater}, {Lindblad}  \&
  {Sandqvist}}{{Hummel} et~al.}{1987}]{hummel2}
{Hummel} E.,  {Jorsater} S.,  {Lindblad} P.~O.,   {Sandqvist} A.,  1987, \aap,
  \href {http://adsabs.harvard.edu/abs/1987A%26A...172...51H} {172, 51}

\bibitem[\protect\citeauthoryear{{Koribalski}, {Staveley-Smith}, {Kilborn},
  {Ryder}  \& {Kraan-Korteweg}}{{Koribalski} et~al.}{2004}]{redshift613}
{Koribalski} B.~S.,  {Staveley-Smith} L.,  {Kilborn} V.~A.,  {Ryder} S.~D.,
  {Kraan-Korteweg} e.~a.,  2004, \mn@doi [\aj] {10.1086/421744}, \href
  {https://ui.adsabs.harvard.edu/abs/2004AJ....128...16K} {128, 16}

\bibitem[\protect\citeauthoryear{{Kormendy}}{{Kormendy}}{1979}]{kormendy}
{Kormendy} J.,  1979, \mn@doi [\apj] {10.1086/156782}, \href
  {http://adsabs.harvard.edu/abs/1979ApJ...227..714K} {227, 714}

\bibitem[\protect\citeauthoryear{{Kormendy} \& {Ho}}{{Kormendy} \&
  {Ho}}{2013}]{reviewcoevolution}
{Kormendy} J.,  {Ho} L.~C.,  2013, \mn@doi [\araa]
  {10.1146/annurev-astro-082708-101811}, \href
  {https://ui.adsabs.harvard.edu/abs/2013ARA&A..51..511K} {51, 511}

\bibitem[\protect\citeauthoryear{{Lucy}}{{Lucy}}{1974}]{lucy}
{Lucy} L.~B.,  1974, \mn@doi [\aj] {10.1086/111605}, \href
  {http://adsabs.harvard.edu/abs/1974AJ.....79..745L} {79, 745}

\bibitem[\protect\citeauthoryear{{Mazzuca}, {Knapen}, {Veilleux}  \&
  {Regan}}{{Mazzuca} et~al.}{2008}]{mazzuca}
{Mazzuca} L.~M.,  {Knapen} J.~H.,  {Veilleux} S.,   {Regan} M.~W.,  2008,
  \mn@doi [The Astrophysical Journal Supplement Series] {10.1086/522338}, \href
  {https://ui.adsabs.harvard.edu/\#abs/2008ApJS..174..337M} {174, 337}

\bibitem[\protect\citeauthoryear{{Menezes}, {Steiner}  \& {Ricci}}{{Menezes}
  et~al.}{2014}]{rob1}
{Menezes} R.~B.,  {Steiner} J.~E.,   {Ricci} T.~V.,  2014, \mn@doi [\mnras]
  {10.1093/mnras/stt2381}, \href
  {http://adsabs.harvard.edu/abs/2014MNRAS.438.2597M} {438, 2597}

\bibitem[\protect\citeauthoryear{{Menezes}, {da Silva}, {Ricci}, {Steiner},
  {May}  \& {Borges}}{{Menezes} et~al.}{2015}]{rob2}
{Menezes} R.~B.,  {da Silva} P.,  {Ricci} T.~V.,  {Steiner} J.~E.,  {May} D.,
  {Borges} B.~W.,  2015, \mn@doi [\mnras] {10.1093/mnras/stv629}, \href
  {http://adsabs.harvard.edu/abs/2015MNRAS.450..369M} {450, 369}

\bibitem[\protect\citeauthoryear{{Menezes}, {Ricci}, {Steiner}, {da Silva},
  {Ferrari}  \& {Borges}}{{Menezes} et~al.}{2019}]{gmostreat}
{Menezes} R.~B.,  {Ricci} T.~V.,  {Steiner} J.~E.,  {da Silva} P.,  {Ferrari}
  F.,   {Borges} B.~W.,  2019, \mn@doi [\mnras] {10.1093/mnras/sty3337}, \href
  {https://ui.adsabs.harvard.edu/\#abs/2019MNRAS.483.3700M} {483, 3700}

\bibitem[\protect\citeauthoryear{{Miyamoto}, {Nakai}, {Seta}, {Salak}, {Nagai}
  \& {Kaneko}}{{Miyamoto} et~al.}{2017}]{miyamoto1}
{Miyamoto} Y.,  {Nakai} N.,  {Seta} M.,  {Salak} D.,  {Nagai} M.,   {Kaneko}
  H.,  2017, \mn@doi [Publications of the Astronomical Society of Japan]
  {10.1093/pasj/psx076}, \href
  {https://ui.adsabs.harvard.edu/\#abs/2017PASJ...69...83M} {69, 83}

\bibitem[\protect\citeauthoryear{{Miyamoto}, {Seta}, {Nakai}, {Watanabe},
  {Salak}  \& {Ishii}}{{Miyamoto} et~al.}{2018}]{miyamoto2}
{Miyamoto} Y.,  {Seta} M.,  {Nakai} N.,  {Watanabe} Y.,  {Salak} D.,   {Ishii}
  S.,  2018, \mn@doi [Publications of the Astronomical Society of Japan]
  {10.1093/pasj/psy016}, \href
  {https://ui.adsabs.harvard.edu/\#abs/2018PASJ...70L...1M} {70, L1}

\bibitem[\protect\citeauthoryear{{Nasonova}, {de Freitas Pacheco}  \&
  {Karachentsev}}{{Nasonova} et~al.}{2011}]{distancia}
{Nasonova} O.~G.,  {de Freitas Pacheco} J.~A.,   {Karachentsev} I.~D.,  2011,
  \mn@doi [\aap] {10.1051/0004-6361/201016004}, \href
  {https://ui.adsabs.harvard.edu/#abs/2011A&A...532A.104N} {532, A104}

\bibitem[\protect\citeauthoryear{{Richardson}}{{Richardson}}{1972}]{rich}
{Richardson} W.~H.,  1972, Journal of the Optical Society of America
  (1917-1983), \href {http://adsabs.harvard.edu/abs/1972JOSA...62...55R} {62,
  55}

\bibitem[\protect\citeauthoryear{{S{\'a}nchez-Bl{\'a}zquez}
  et~al.,}{{S{\'a}nchez-Bl{\'a}zquez} et~al.}{2006}]{blazquez}
{S{\'a}nchez-Bl{\'a}zquez} P.,  et~al., 2006, \mn@doi [\mnras]
  {10.1111/j.1365-2966.2006.10699.x}, \href
  {http://adsabs.harvard.edu/abs/2006MNRAS.371..703S} {371, 703}

\bibitem[\protect\citeauthoryear{{Sargent}, {Schechter}, {Boksenberg}  \&
  {Shortridge}}{{Sargent} et~al.}{1977}]{sargent}
{Sargent} W.~L.~W.,  {Schechter} P.~L.,  {Boksenberg} A.,   {Shortridge} K.,
  1977, \mn@doi [\apj] {10.1086/155052}, \href
  {https://ui.adsabs.harvard.edu/abs/1977ApJ...212..326S} {212, 326}

\bibitem[\protect\citeauthoryear{{Schechter}}{{Schechter}}{1983}]{schechter}
{Schechter} P.~L.,  1983, \mn@doi [The Astrophysical Journal Supplement Series]
  {10.1086/190875}, \href
  {https://ui.adsabs.harvard.edu/#abs/1983ApJS...52..425S} {52, 425}

\bibitem[\protect\citeauthoryear{{Storchi-Bergmann} \&
  {Schnorr-M{\"u}ller}}{{Storchi-Bergmann} \&
  {Schnorr-M{\"u}ller}}{2019}]{reviewfeeeding}
{Storchi-Bergmann} T.,  {Schnorr-M{\"u}ller} A.,  2019, \mn@doi [Nature
  Astronomy] {10.1038/s41550-018-0611-0}, \href
  {https://ui.adsabs.harvard.edu/abs/2019NatAs...3...48S} {3, 48}

\bibitem[\protect\citeauthoryear{{Winge}, {Riffel}  \&
  {Storchi-Bergmann}}{{Winge} et~al.}{2009}]{starlightinfrared}
{Winge} C.,  {Riffel} R.~A.,   {Storchi-Bergmann} T.,  2009, \mn@doi [\apjs]
  {10.1088/0067-0049/185/1/186}, \href
  {http://adsabs.harvard.edu/abs/2009ApJS..185..186W} {185, 186}

\bibitem[\protect\citeauthoryear{{da Silva}, {Menezes}  \& {Steiner}}{{da
  Silva} et~al.}{2020}]{paty}
{da Silva} P.,  {Menezes} R.~B.,   {Steiner} J.~E.,  2020, \mn@doi [\mnras]
  {10.1093/mnras/staa007}, \href
  {https://ui.adsabs.harvard.edu/abs/2020MNRAS.492.5121D} {492, 5121}

\makeatother
\end{thebibliography}

%%%%%%%%%%%%%%%%%%%%%%%%%%%%%%%%%%%%%%%%%%%%%%%%%%

%%%%%%%%%%%%%%%%% APPENDICES %%%%%%%%%%%%%%%%%%%%%

\appendix

\section{Spectral Synthesis on SIFS data cube}\label{sifs_starlighttotal}

The SIFS data cube of NGC~613 has a wide FOV with which it is possible to study both nuclear and circumnuclear structures, such as the circumnuclear star-forming ring (see section \ref{starlight_regioesHII}). In order to see the general results of this FOV, we applied the spectral synthesis with the \textsc{starlight} software (see section \ref{starlight_def} for more details) to the entire FOV of this data cube. 

\begin{figure}
\begin{center}
   \includegraphics[scale=0.5]{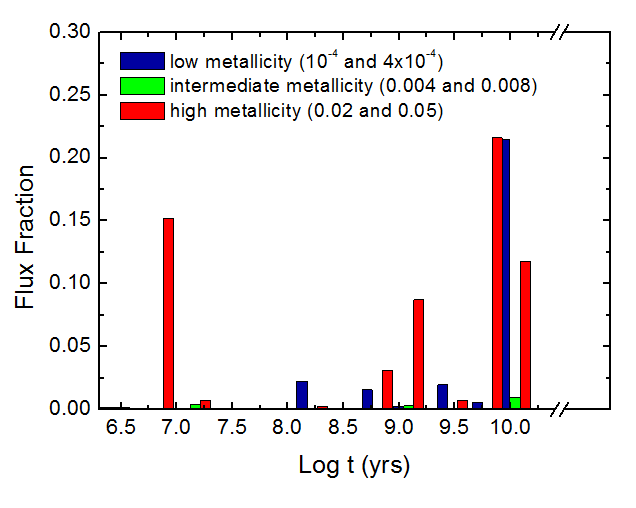}
  \caption{Histogram of the flux fractions of the stellar populations detected from the spectral synthesis applied to the entire FOV of the SIFS data cube. As in the GMOS data cube results, we did not find any flux fraction associated with the power law with 1.5 index representing a featureless continuum emission of the AGN, which should be after the break on the histogram.  \label{histogramasifstotal}}
\end{center}
\end{figure}

From the results of the spectral synthesis applied to the entire data cube, we created a histogram of the flux fraction of the detected stellar populations, separated by age and metallicity, as shown in Fig.~\ref{histogramasifstotal}. We notice that there are, mainly, four star-forming events, two around 10~Gyr, one around 1~Gyr and one $\sim$~10~Myr. There is a general predominance of high-metallicity stars (0.02 and 0.05). However, there is a considerable flux fraction associated with low-metallicity (10$^{-4}$ and $4~\times~10^{-4}$) old stellar populations ($\sim$~10~Gyr). 

\begin{figure}
\begin{center}
   \includegraphics[scale=0.35]{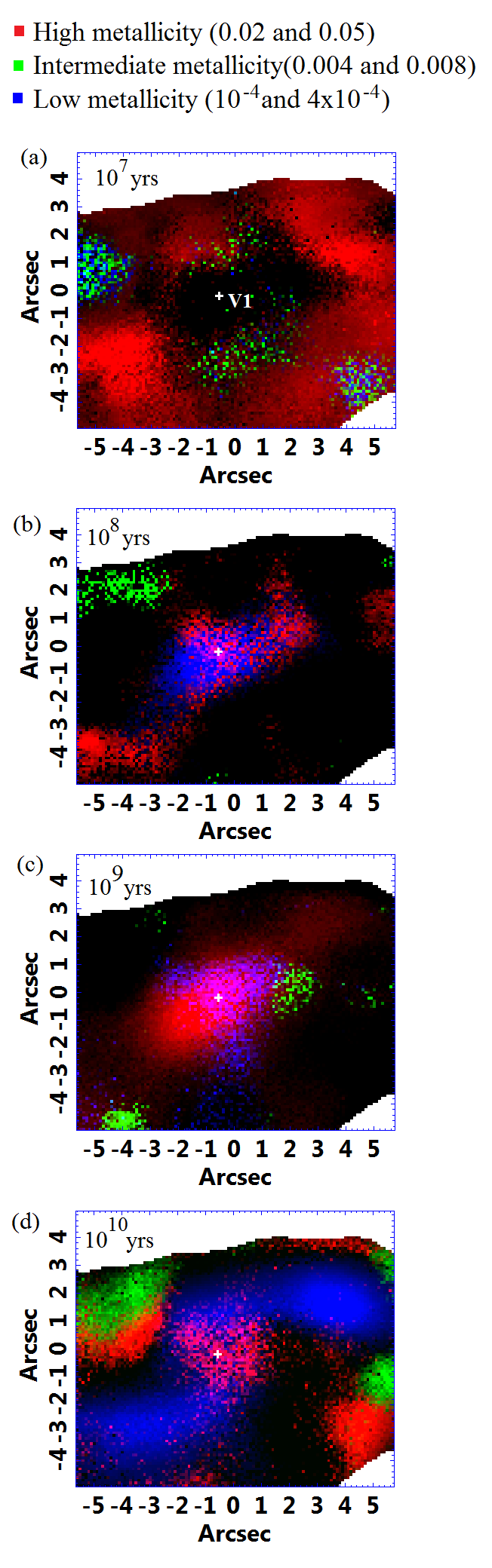}
\caption{Maps of the flux fractions of the stellar populations obtained from the spectral synthesis applied to the SIFS data cube: (panel a) 10$^6$, (panel b) 10$^8$, (panel c) 10$^9$ and (panel d) 10$^{10}$~yr. The blue colour represents low-metallicity stellar populations (10$^{-4}$ and 4~$\times~10^{-4}$), the green colour represents intermediate-metallicity stellar populations (0.004 e 0.008), and the red colour represents the high-metallicity stellar populations (0.02 and 0.05). The white cross represents the V1 position and its size represents the 3$\sigma$ uncertainty. \label{mapastarlightsifs}}
\end{center}
\end{figure}

As the spectral synthesis was applied to each spectrum of the data cube, it was possible to create maps of the flux due to each stellar population taken into account (see Fig.~\ref{mapastarlightsifs}). We notice that there is a concentration of young stellar populations ($\sim$~10$^7$~yr) with high metallicity in the circumnuclear regions, not only in the circumnuclear ring (Fig.~\ref{mapastarlightsifs}a). There is a concentration of low and high metallicities stellar populations with ages between 10$^8$ and 10$^9$~yr in the centre of the FOV (Figs. \ref{mapastarlightsifs}b and c). We notice a structure that is similar to two arms coming from the centre of the FOV, whose age is $\sim$~10~Gyr with low-metallicity stars. 

\begin{figure}
\begin{center}
   \includegraphics[scale=0.35]{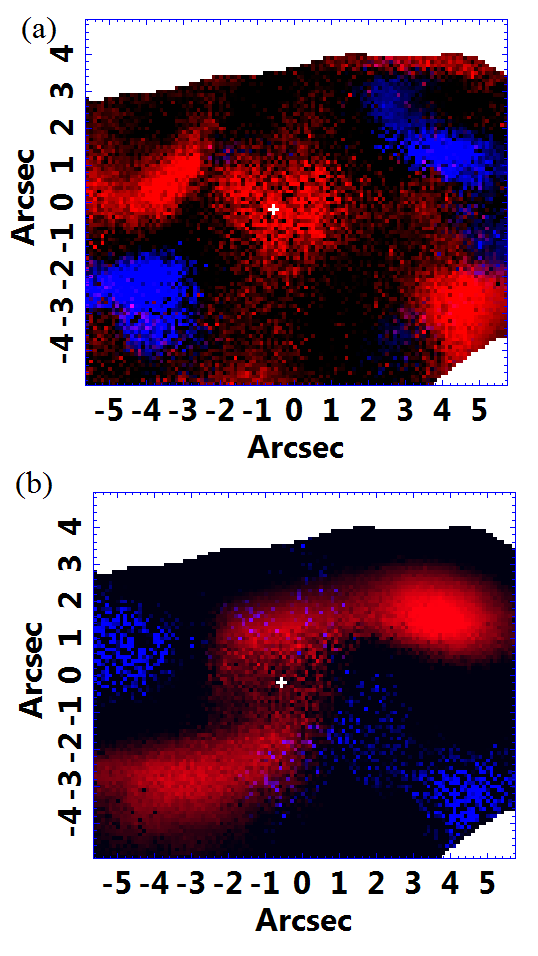}
\caption{Maps of the flux fractions obtained with the spectral synthesis applied to the SIFS data cube of the stellar populations: (panel a) with high (0.02 and 0.05) and (panel b) with low metallicity (10$^{-4}$ and 4~$\times~10^{-4}$). The red colour represents the old stars with $\sim$ 10$^{10}$ yr and the blue colour represents the young stars with $\sim$~10$^{7}$ yr. The white cross represents the position of V1 and its size represents 3$\sigma$ uncertainty.  \label{mapastarlightsifs_anel}}
\end{center}
\end{figure}

Fig.~\ref{mapastarlightsifs_anel} shows, mainly, the structures located at the circumnuclear ring of NGC~613. The map of Fig.~\ref{mapastarlightsifs_anel}(a) indicates the high-metallicity stars. We see that the old stellar populations (in red) are located at regions opposite to the young stellar populations (in blue). We also see that, by comparing this image with Fig.~17(b) of Paper~I, the areas of young stellar formation of this image are close to the two points where the ring meets the bar. The situation is inverse when we look to the low-metallicity stellar populations map (Fig.~\ref{mapastarlightsifs_anel}b). Again, we can see that young and old stellar populations are located at opposite regions. However, whereas, in the image of high metallicity, there were young stars, in this image of low metallicity, there are old stars, and vice versa. Such result is compatible with the ''pearls on string'' scenario \citep{BOKERIAU,boker,falcon613}, in which stellar populations are formed in the two points where the ring is being fed by the bar and, due to the rotation of the ring, as the star-forming regions rotate, the stellar populations get old and then the old stars are located at opposite sides from the young stellar populations. In the low-metallicity results, the inversion of those positions relative to the high-metallicity image can be explained by the fact that those star-forming regions (with low metallicity) might be turned around in the ring more than the high-metallicity stars.

\begin{figure}
\begin{center}
   \includegraphics[scale=0.35]{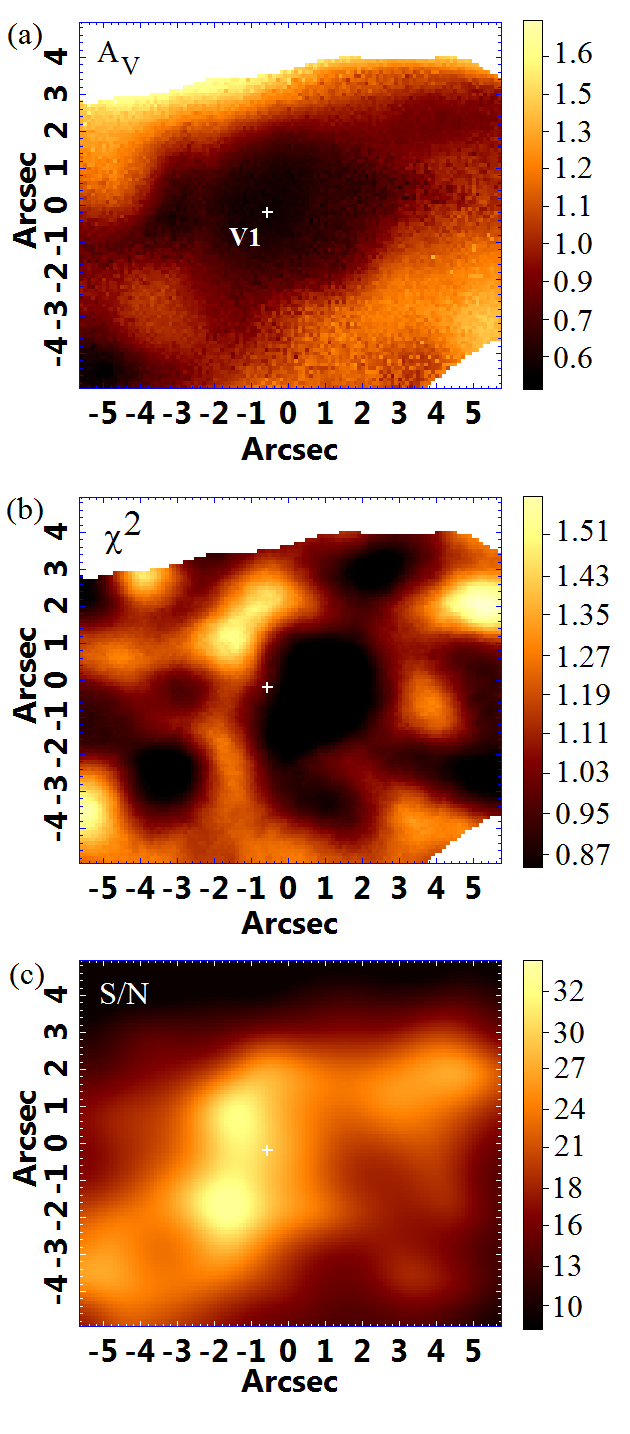}
  \caption{(Panel a): map of the extinction obtained from the spectral synthesis applied to SIFS data cube with \textsc{starlight}; (panel b): map of $\chi^2$ of the fits of this method; and (panel c): map of the S/N obtained from the spectral synthesis with the pPXF method. This last map was used to mask the regions of the SIFS data cube maps used in the results with S/N < 10. The white cross represents the V1 position and its size represents the 3$\sigma$ uncertainty. \label{avqui2sn}}
\end{center}
\end{figure}

The A$_v$ map (Fig.~\ref{avqui2sn}a) shows that most of the extinction comes from the circumnuclear regions, since in the centre the values are lower. The $\chi^2$ map (Fig.~\ref{avqui2sn}b) shows that the \textsc{starlight} fits well succeeded in the areas of the FOV where the results were considered. Such areas were determined by the map of S/N of Fig.~\ref{avqui2sn}(c), obtained with pPXF, that shows that the superior and the left lower FOV edge have values lower than 10. Thus, those regions were removed from the results due to high uncertainties.

%%%%%%%%%%%%%%%%%%%%%%%%%%%%%%%%%%%%%%%%%%%%%%%%%%

% Don't change these lines
\bsp	% typesetting comment
\label{lastpage}
\end{document}